\documentclass[11pt,a4paper]{article}
\pdfoutput=1

\usepackage[toc,page]{appendix}
 \usepackage{comment}
\usepackage[english]{babel}
\usepackage{caption}
\usepackage{subcaption}
\usepackage{amsfonts}

\usepackage{color}
\usepackage[pdftex]{graphicx}
\usepackage{bm}
\usepackage{graphicx}
\usepackage{amsmath}

\usepackage{color,soul}
 \usepackage{hyperref}
 
 \usepackage{pdfpages}
\newcommand{\fig}[1]{Fig.\ \ref{#1}}

\begin{document}
\begin{titlepage}
\thispagestyle{empty}

\vspace{2cm}
\begin{center}
\font\titlerm=cmr10 scaled\magstep4
\font\titlei=cmmi10
scaled\magstep4 \font\titleis=cmmi7 scaled\magstep4 {\Large{\textbf{QGP probes from a dynamical holographic model of AdS/QCD}
\\}}
\setcounter{footnote}{0}
\vspace{1.5cm} \noindent{{%\large
 S. Heshmatian$^{a}$\footnote{e-mail:heshmatian@bzte.ac.ir } and R. Morad$^{b,c}$\footnote{e-mail:rmorad@tlabs.ac.za}
}}\\
\vspace{0.8cm}

{\it $^{a}$ Department of Engineering Sciences and Physics, Buein Zahra Technical University, Buein Zahra, Qazvin, Iran\\}
{\it $^{b}$ UNESCO UNISA ITL/NRF Africa Chair in Nanosciences/Nanotechnology, College of Graduate Studies, University of South Africa (UNISA), South Africa \\}
{\it $^{c}$ Nanosciences African Network, Materials Research Department, iThemba LABS, National Research Foundation, South Africa}

\vspace*{.4cm}
\end{center}
\vskip 2em
\setcounter{footnote}{0}
\begin{abstract}
In this paper, we employ the gauge/gravity duality to study some features of the quark-gluon plasma. For this purpose, we implement a holographic QCD model constructed from an Einstein-Maxwell-Dilaton gravity at finite temperature and finite chemical potential. The model captures both the confinement and deconfinement phases of QCD and we use it to study the effect of temperature and chemical potential on a heavy quark moving through the plasma. We calculate the drag force, Langevin diffusion coefficients and also the jet quenching parameter, and our results align with other holographic QCD models and the experimental data.

\end{abstract}
\end{titlepage}
\tableofcontents

\section{Introduction}

Quark-gluon plasma (QGP) produced at the Relativistic Heavy Ion Collider (RHIC) and the Large Hadron Collider (LHC) is a strongly coupled plasma whose dynamics is dominated by non-perturbative effects \cite{Baier:1996kr,Eskola:2004cr}. Since the perturbative quantum chromodynamics (QCD) is applicable only in the weak coupling regime, the lattice QCD techniques could be hired for understanding the static equilibrium thermodynamics of such matter. Alternatively, the "AdS/CFT correspondence" \cite{Maldacena:1997re}-\cite{CasalderreySolana:2011us} provides a non-perturbative tool to examine the dynamical quantities of the strongly coupled QGP. The "AdS/CFT correspondence" indicates a duality between the $\mathcal{N} = 4\, SU(N_c)$ super-Yang-Mills theory and type IIB string theory on $AdS_5 \times S^5$, which is a powerful tool to study the strongly coupled gauge theory in the large $N_c$ limit and large ’t Hooft coupling. The original duality maps an asymptotically AdS space to a conformal gauge theory at zero temperature. On the other hand, due to the fact that the  physical quantities of the quark-gluon plasma are temperature dependent, many attempts were made to extend the original duality to holographic models describing the QGP at finite temperature using top-down \cite{Polchinski:2000uf}-\cite{Sakai:2004cn} or bottom-up models \cite{He:2013qq}-\cite{Panero:2009tv}.\\

One of the interesting features of QCD is the confinement-deconfinement crossover where the QCD coupling constant becomes very large. This phenomenon leads to a strong suppression of quarkonium near the crossover temperature, as the confined phase of QCD is at a lower temperature and density compared to the deconfined phase \cite{Adare}-\cite{Matsui:1986dk}. Lattice QCD calculations demonstrate that the entropy of the quark-antiquark pair develops a peak around the crossover area \cite{Kaczmarek:2005ui,Hashimoto:2014fha} which indicates the strong interaction between the quark-antiquark pair in this region. The quark-antiquark entropy in both phases and the confinement-deconfinement crossover have both been thoroughly studied utilizing the AdS/CFT correspondence and generalized dual holographic models. \cite{Maldacena:1998im}-\cite{Zhang:2016fwr}. \\

Among holographic models of QCD, the model constructed from a black hole solution using the Einstein-Maxwell-Dilaton gravity is beneficial as it incorporates the temperature dependency of the entropy of the quark-antiquark pair in both confined and deconfined phases \cite{Dudal:2018ztm,Dudal:2017max}. On the other hand, this holographic model includes chemical potential as the QCD equation of state and phase diagram depends on the chemical potential. On the gravity side, increasing temperature leads to a phase transition from thermal AdS to a black hole using a specific choice of an arbitrary function $A(z)$. In the boundary theory, these are dual to phase transitions from confined to deconfined phases. The results of this model for the quark-antiquark entropy are in good agreement with those from Lattice QCD and therefore, this holographic model could be a capable model to study the QCD feature in confined and deconfined phases. The gravity solution was then extended with a magnetic field to analyzed the quark-antiquark free energy and entropy from holographic point of view \cite{Bohra:2019ebj},\cite{Jena:2022nzw}.\\  

Jets of quarks propagating through the medium are the most interesting objects produced in QGP and their interaction with the medium is one of the challenging problems in new physics. The transport coefficient, also known as the jet quenching parameter, is one of the most interesting experimental observables associated with quark energy loss in the hot medium observed at RHIC and LHC \cite{Adams:2005dq}-\cite{Burke:2013yra}. It is defined as the average squared transverse momentum transferred from a traversing parton to the medium, per unit mean free path \cite{DEramo:2010wup}. The parton energy loss through the medium can also be investigated using the drag force experienced by heavy quarks moving in the medium. Due to the strong nature of the interaction, holographic models of QCD are utilized as powerful tools to investigate the dynamics of this phenomena and characterize properties of jets and QGP medium as well as their interactions which leads to the quark energy loss. Also, heavy quarks moving in QGP undergo a Brownian-like motion which could be studied by the Langevin diffusion coefficients \cite{Rapp:2009my},\cite{Dunkel:2008ngc}.\\

According to the AdS/CFT correspondence, a heavy quark is illustrated as a fundamental string attached to a flavor brane. The string endpoint could be considered as a quark in the boundary theory and the string itself can be considered as a gluonic cloud surrounding the quark. The drag force of a quark moving in the medium can be measured from the corresponding momentum flowing of an open trailing string to the bulk \cite{Domurcukgul:2021qfe}-\cite{Gubser:2006bz}. The quantum fluctuations of this trailing string are also dual to the momentum broadening of a heavy quark moving in the QGP and can be computed along the quark direction of the motion as well as the transverse plane. The stochastic Brownian motion is associated with the correlators of the string fluctuations and can be expressed in terms of the Langevin coefficients \cite{Xiong:2019wik}-\cite{Zhu:2020wds}. According to this duality, the jet quenching parameter is related to the thermal expectation value of the light-like Wilson loop operator created from the trajectory of two string endpoints, and many attempts were made to compute this parameter in the context of the AdS/CFT correspondence \cite{Liu:2006ug}-\cite{Zakharov:1997uu}. The holographic energy loss of a quark moving in a strongly coupled QGP has been studied in \cite{Giataganas:2012zy} for a strongly coupled anisotropic $\mathcal{N}=4$ super Yang-Mills plasma and also, in \cite{Zhou:2022izh} using another version of the Einstein-Maxwell-dilaton model where the anisotropy is introduced at one spatial direction in metric \cite{Arefeva:2020byn}.\\

In this paper, we study the QGP aspects such as the drag force, Langevin coefficients, and the jet quenching parameter for a heavy quark moving through the plasma using the gauge/gravity duality. For this purpose, we implement the dynamical holographic QCD model of ref.\cite{Dudal:2017max} as the dual boundary theory involves temperature in both confined and deconfined phases as well as the chemical potential and has a qualitative agreement with Lattice results for the quark-antiquark entropy in both confined and deconfined phases, therefore it would be an appropriate model to examine the QGP features.\\

This paper is organized as follows: In section \ref{EMDG} we briefly review the holographic QCD model constructed from the Eintein-Maxwell-Dilaton gravity introduced in \cite{Dudal:2017max}. In section \ref{dragforce} we discuss the drag force on heavy quark moving in the dynamical holographic model of QCD. In section \ref{Langevin}, the Langevin coefficients are calculated. In section \ref{JetQ}, the jet quenching parameter is studied, and in section \ref{conclusion}, a summary and conclusion are presented. 

%%%%%%%%%%%%%%%%%%%%%%%%%%%%%%%%%%%%%%%%%%%%%%%%%%%%%%%%%%%%%%%%%%%%%%%%%
\section{Einstein-Maxwell-dilaton gravity}
\label{EMDG}

In this section, we review the Einstein-Maxwell-dilaton gravity (EMD) model at finite and zero temperature \cite{Dudal:2018ztm,Dudal:2017max} which is described by the following action in 5 dimensions,
\begin{eqnarray}
&&S_{EM} =  -\frac{1}{16 \pi G_5} \int \mathrm{d^5}x \sqrt{-g} \ \ \bigl[R-\frac{f(\phi)}{4}F_{MN}F^{MN} -\frac{1}{2}\partial_{M}\phi \partial^{M}\phi -V(\phi)\bigr] \,.
\label{actionEF}
\end{eqnarray}
where $G_5$ is the Newton constant, $V(\phi)$ is the potential of the dilaton field and $f(\phi)$ is the gauge kinetic function representing the coupling between the dilaton and the gauge field $A_{M}$. 

By assuming the following ansatz for the metric, gauge field, and dilaton field,
\begin{eqnarray}
& & ds^2=\frac{L^2 e^{2 A(z)}}{z^2}\biggl(-g(z)dt^2  + dx_1^2+dx_2^2+dx_3^2 + \frac{dz^2}{g(z)}\biggr)\,, \nonumber \\
& & A_{M}=A_{t}(z), \ \ \ \ \phi=\phi(z),
\label{metric}
\end{eqnarray}
where L is the AdS length scale and fields are assumed to be functions of extra radial coordinate $z$, one can solve the equations of motion analytically in terms of $A(z)$ and $f(z)$ \cite{He:2013qq,Yang:2015aia}
\begin{eqnarray}
&&g(z)=1-\frac{\int_{0}^{z} dx \  x^3 e^{-3A(x)} \int_{x_c}^{x} dx_1 \  \frac{x_1 e^{-A(x_1)}}{f(x_1)}}{\int_{0}^{z_h} dx \ x^3 e^{-3A(x)}  \int_{x_c}^{x} dx_1 \  \frac{x_1 e^{-A(x_1)}}{f(x_1)} }, \nonumber \\
&&\phi'(z)=\sqrt{6(A'^2-A''-2 A'/z)}, \nonumber \\
&& A_{t}(z)=\sqrt{\frac{-1}{\int_{0}^{z_h} dx \ x^3 e^{-3A(x)}  \int_{x_c}^{x} dx_1 \  \frac{x_1 e^{-A(x_1)}}{f(x_1)}}}   \int_{z_h}^{z} dx \  \frac{x e^{-A(x)}}{f(x)}, \nonumber \\
&&V(z)=-\frac{3z^2ge^{-2A}}{L^2}\bigl[A''+A' \bigl(3A'-\frac{6}{z}+\frac{3g'}{2g}\bigr)-\frac{1}{z}\bigl(-\frac{4}{z}+\frac{3g'}{2g}\bigr)+\frac{g''}{6g} \bigr]\,.
\label{metsolution}
\end{eqnarray}
The spacetime asymptotic boundary is located at $z=0$ where $g(z)$ goes to 1, and $z=z_h$ is the position of the horizon with the boundary condition of $g(z_h)=0$.
It is shown that by considering the simple form of $f(z)=e^{c z^2 -A(z)}$, one can reproduce the linear Regge trajectory of the discrete spectrum of the mesons. The value of $c=1.16 \ \text{GeV}^2$ is fixed by matching the holographic meson mass spectrum to the lowest-lying heavy meson states \cite{He:2013qq}.\\
The gravity solution corresponds to a black hole with horizon located at $z_h$, is then given by
\begin{eqnarray}
&&g(z)=1-\frac{1}{\int_{0}^{z_h} dx \ x^3 e^{-3A(x)}} \biggl[\int_{0}^{z} dx \ x^3 e^{-3A(x)} + \frac{2 c \mu^2}{(1-e^{-c z_{h}^2})^2} \det \mathcal{G}  \biggr],\nonumber \\
&&\phi'(z)=\sqrt{6(A'^2-A''-2 A'/z)}, \nonumber \\
&& A_{t}(z)=\mu \frac{e^{-c z^2}-e^{-c z_{h}^2}}{1-e^{-c z_{h}^2}}, \nonumber \\
&&V(z)=-\frac{3z^2ge^{-2A}}{L^2}\left[A''+A' \left(3A'-\frac{6}{z}+\frac{3g'}{2g}\right)-\frac{1}{z}\left(-\frac{4}{z}+\frac{3g'}{2g}\right)+\frac{g''}{6g} \right] \,,
\label{metsolution1}
\end{eqnarray}
where $\mu$ is the chemical potential and
\[
\det \mathcal{G} =
\begin{vmatrix}
\int_{0}^{z_h} dx \ x^3 e^{-3A(x)} & \int_{0}^{z_h} dx \ x^3 e^{-3A(x)- c x^2} \\
\int_{z_h}^{z} dx \ x^3 e^{-3A(x)} & \int_{z_h}^{z} dx \ x^3 e^{-3A(x)- c x^2}
\end{vmatrix}\,.
\]
The Hawking temperature and entropy of this solution are 
\begin{eqnarray}
T&=& \frac{z_{h}^3 e^{-3 A(z_h)}}{4 \pi \int_{0}^{z_h} dx \ x^3 e^{-3A(x)}} \biggl[ 1+\frac{2 c \mu^2 \bigl(e^{-c z_h^{2}}\int_{0}^{z_h} dx \ x^3 e^{-3A(x)}-\int_{0}^{z_h} dx \ x^3 e^{-3A(x)}e^{-c x^{2}} \bigr)}{(1-e^{-c z_h^{2}})^2} \biggr] \nonumber \\
S_{BH}&=& \frac{L^3 e^{3 A(z_h)}}{4 G_5 z_{h}^3}\, .
%\,\,\,\, ,\,\,\,\,\,  S_{SYM}\,=\,\frac{\pi^3\, L^3 \,T^3}{4 G_5}
\label{Htemp}
\end{eqnarray}

There is another solution for the Einstein-Maxwell-dilaton equations which corresponds to a thermal-AdS space. This solution is obtained by taking the $z_h\rightarrow\infty$ limit of the solution \ref{metsolution1} which results in $g(z)=1$. The chemical potential goes to zero in this limit and therefore, the thermal gas solution has zero chemical potential. This thermal solution is asymptotically AdS. Nevertheless, it can have a non-trivial structure in the bulk due to the scale factor $A(z)$.

It should be highlighted that, traditionally, one would construct a gravitational background by fixing $V(\phi)$ and solve the Einstein equations. Instead, for the model summarized here, the background is constructed by fixing $A(z)$ which leads to an implicit dependency on $T$ and $\mu$ for the dilaton potential. Consequently, the thermodynamic laws are slightly violated. This is a known issue for this construction method, however, it has been verified that the dependency of dilaton potential on these parameters is generally minor. For more detail please refer to \cite{Bohra:2019ebj}.

\begin{figure}
\begin{subfigure}{.45\textwidth}
%%%%%%%%%%
  \centering
  \includegraphics[width=0.95\linewidth]{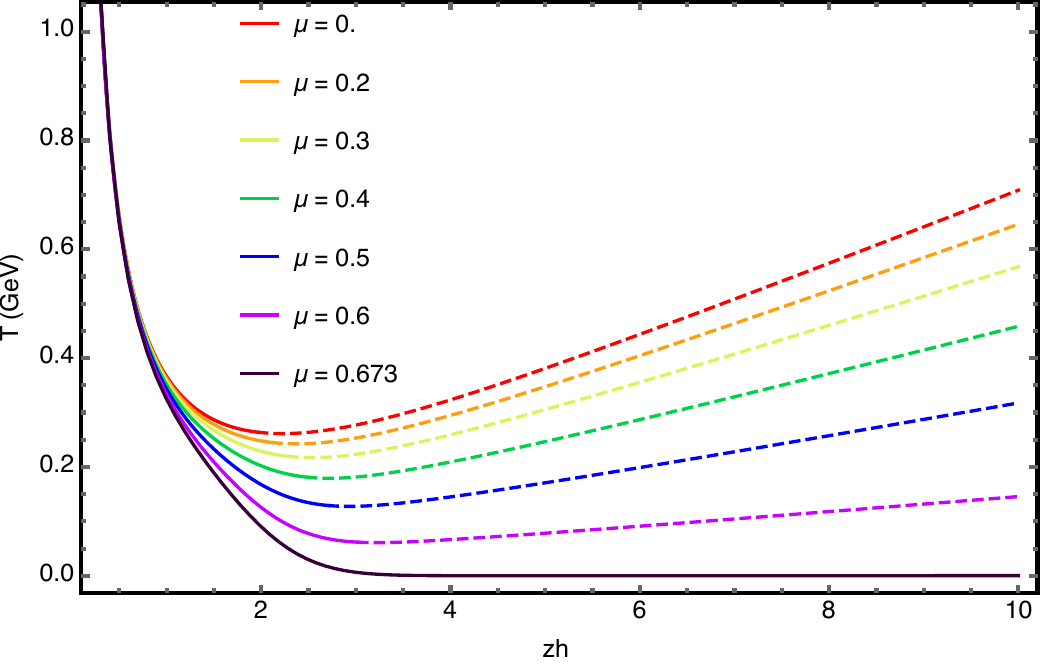}
   \caption{}
  \label{temperature}
\end{subfigure}
%%%%%%%%%%
\begin{subfigure}{.45\textwidth}
  \centering
  \includegraphics[width=0.95\linewidth]{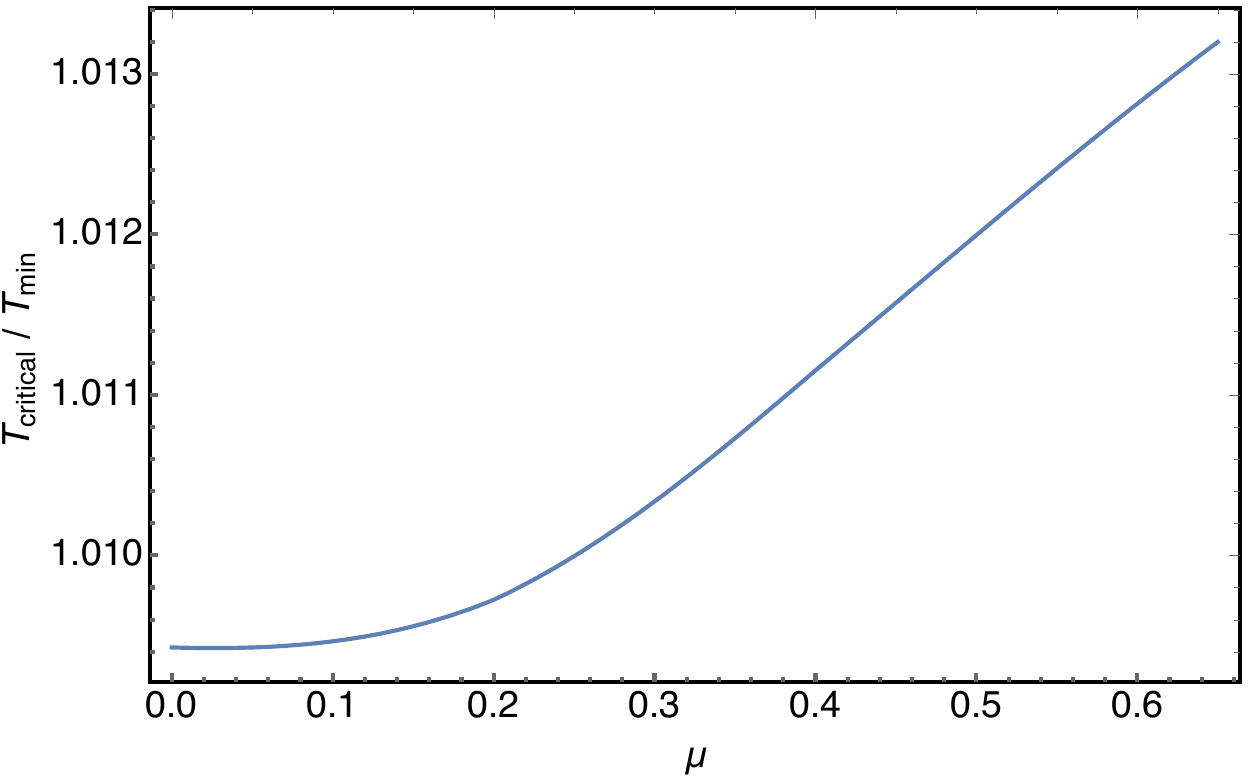}
   \caption{}
  \label{TcTm}
\end{subfigure}
%%%%%%%%%%
\caption{\small (Color online) (\subref{temperature}) The temperature of the solution versus $z_h$ for different values of $\mu$.(\subref{TcTm}) The ratio of critical temperature to the minimum temperature for different values of $\mu$. Stable solutions are represented by solid lines while dashed lines characterize the unstable solutions. }
\label{T}
\end{figure}

\begin{figure}
\begin{subfigure}{.45\textwidth}
%%%%%%%%%%
  \centering
  \includegraphics[width=0.95\linewidth]{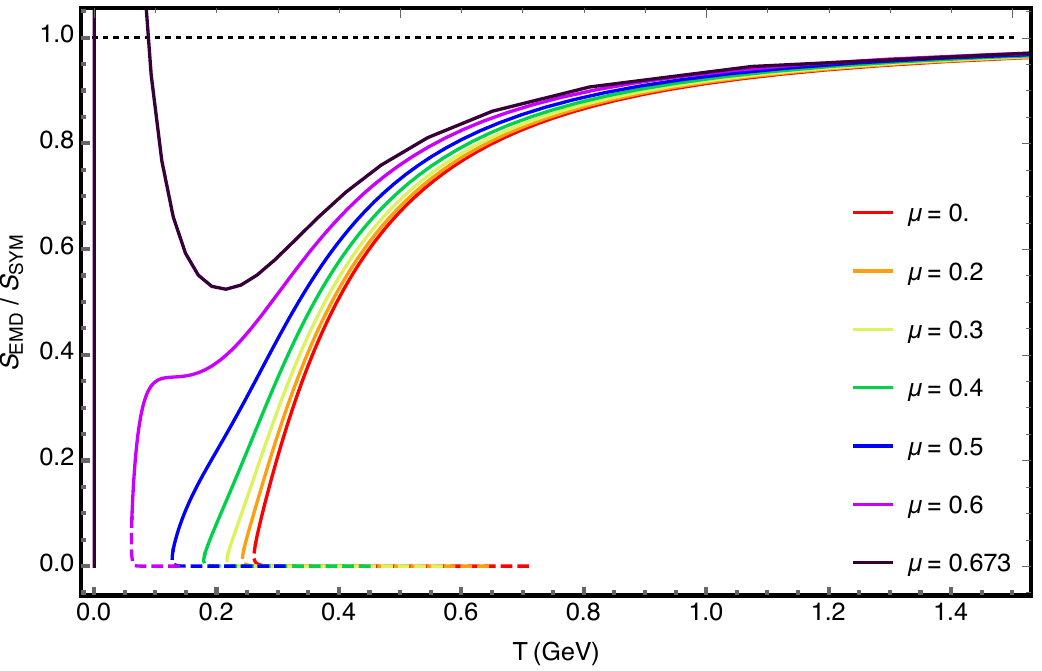}
  \caption{}
  \label{entropyratio}
\end{subfigure}
%%%%%%%%%%
\begin{subfigure}{.45\textwidth}
  \centering
  \includegraphics[width=0.95\linewidth]{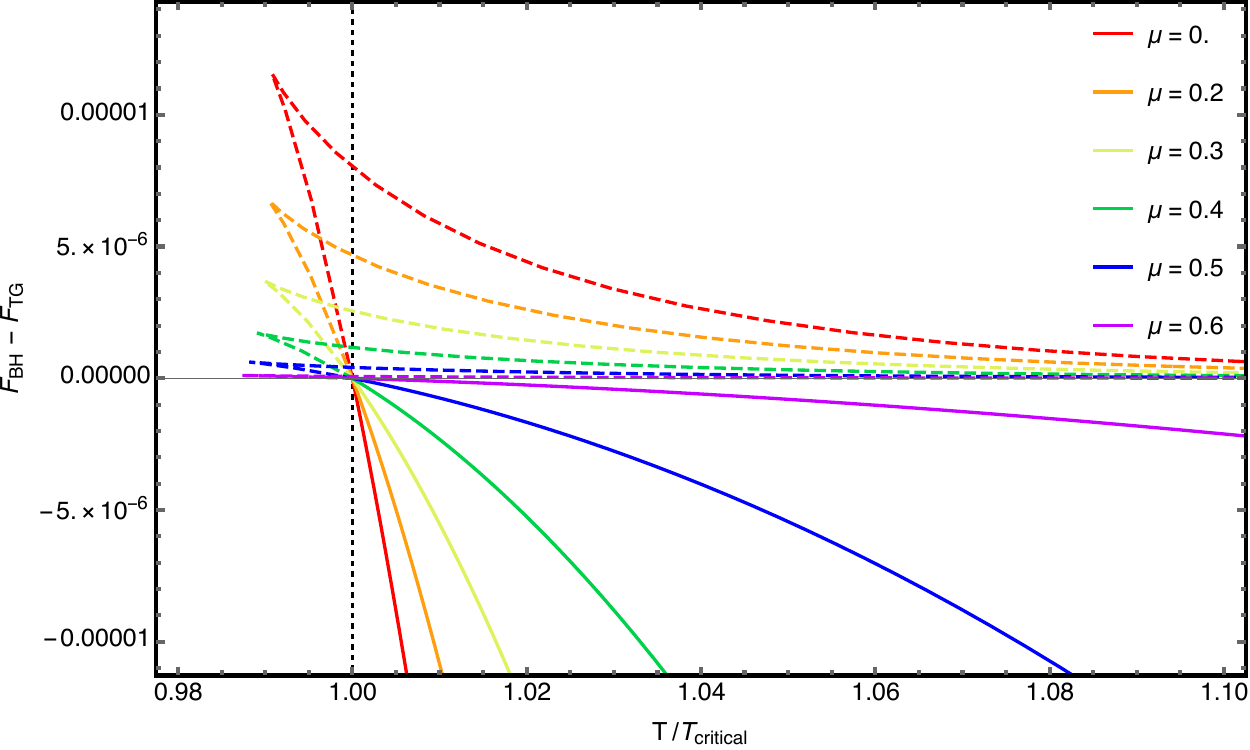}
   \caption{}
  \label{freeenergy}
\end{subfigure}
%%%%%%%%%%
\caption{\small (Color online)(\subref{entropyratio})  The ratio of black hole entropy to the AdS Schwarzschild black hole entropy versus temperature for different values of $\mu$. (\subref{freeenergy}) The free energy versus temperature for different values of chemical potential between zero (red) and $\mu=\mu_c$ (purple). For each value of $\mu$, there is a phase transition between the AdS-BH and thermal gas solution. Here, $\mu_c\,=\,0.673\, GeV$ is the critical value for $\mu$ for which there is no phase transition and the large black hole solution exists from zero temperature to high temperature. Solid (Dashed) lines show the stable (unstable) solutions. }
\label{thermodynamics}
\end{figure}

For the holographic model to describe the confinement-deconfinement phases in the boundary theory, the following form of $A(z)$ is being considered 
\begin{eqnarray}
A(z)=A_2(z)=-\bar{a} z^2\,,
\label{Az}
\end{eqnarray}
which vanishes at the boundary. The value of $\bar{a}=c/8\simeq 0.145$ is determined such that the Hawking/Page phase transition occurs at $T\simeq 270 MeV$ at zero chemical potential. With this choice of $A(z)$, the thermodynamic properties of the solution are shown in Figs.\ref{T} and \ref{thermodynamics}.

In Fig. \ref{temperature}, the temperature of the solution is plotted in terms of $z_h$. At each temperature, for small values of $\mu$, there are two solutions: large black hole (small $z_h$) and small black hole (large $z_h$). The small black hole solution is thermodynamically unstable. Also the black hole solution only exist above a minimum temperature, $T_{min}$ indicating there is a first order phase transition from black hole to thermal-gas phase. Although, this transition happens at some critical temperature $T_c$ above the $T_{min}$. In Fig. \ref{TcTm}, the ratio of critical temperature to the minimum temperature is plotted for different values of $\mu$. This ratio is always larger than one and enhanced by increasing the chemical potential.
Increasing the value of chemical potential in Fig. \ref{temperature}, the branch with negative slope (the small black hole solution) becomes smaller and completely vanishes at some critical chemical potential $\mu=\mu_c=0.673$. At $\mu>\mu_c$, we have a single black hole solution which remains stable at all temperatures. \\

In Fig. \ref{entropyratio}, the normalized entropy of the system is plotted in terms of temperature. By increasing temperature, this ratio reaches the value of one. In Fig. \ref{freeenergy}, the free energy of the solution is plotted in terms of the temperature. The free energy is normalized such that the free energy of thermal-AdS is zero. Therefore, changing the sign of free energy means a first-order phase transition from AdS black hole to thermal-AdS as the temperature decreases which takes place at $T=T_c$. This is the famous Hawking–Page phase transition. In \cite{Dudal:2017max}, this Hawking-Page phase transition on the gravity side is used to determine the confinement/deconfinement phase transition on the dual boundary side. They used the phase for which the free energy of the probe quark-antiquark pair varies linearly with respect to their separation length, as confinement.
Here, we explore this confinement/deconfinement phase transition using the external probes on the gravity side.
In order to study the holographic probes we are interested in, it is more convenient to use the string frame metric which can be obtained using the following standard transformations,
\begin{eqnarray}
& & (g_s)_{MN}=e^{\sqrt{\frac{2}{3}}\phi} g_{MN}, \nonumber \\
& & ds_{s}^2= \frac{L^2 e^{2 A_{s}(z)}}{z^2}\biggl(-g(z)dt^2 + \frac{dz^2}{g(z)} + dx_{1}^2+dx_{3}^2+dx_{3}^2 \biggr)\, ,
\label{stringmetric}
\end{eqnarray}
where $A_{s}(z)=A(z)+\sqrt{\frac{1}{6}} \phi(z)$.
The open string tension $T_s$ in units of the AdS length scale $L^2$ is considered to be $T_s L^2 \simeq 0.1$ \cite{Dudal:2017max}, which is obtained by comparing the numerical results with the lattice QCD estimate of the string tension, i.e. $\sigma_{s}\approx 1/(2.34)^2 GeV^2$  \cite{Sommer:1993ce}. 

%%%%%%%%%%%%%%%%%%%%%%%%%%%%%%%%%%%%%%%%%%%%%%%%%%%%%%%%%%%%%%
\section{Drag Force
\label{dragforce}}
%%%%%%%%%%%%%%%%%%%%%%%%%%%%%%%%%%%%%%%%%%%%%%%%%%%%%%%%%%%%%%

In this section, we consider a heavy quark moves with a constant velocity in one of the spatial directions denoted by x. Therefore, the embedding function of a heavy quark in the static gauge $\tau=t\,,\sigma=z$ is  $X = \{t, x(t,z), z\}$. The action of a fundamental string is given by the Nambu- Goto action as,
\begin{eqnarray}\label{NGaction}
S &=& -\frac{1}{2\pi \alpha'} \int\,d\sigma \,d\tau\,\sqrt{-\gamma}\, \nonumber \\
&=& -\frac{1}{2\pi \alpha'} \int\,d\sigma \,d\tau\,\sqrt{-G_{tt}G_{zz}-G_{tt} G_{xx} x^{\prime \,2} -G_{xx}G_{zz}\dot{x}^2 }\, ,
\end{eqnarray}
where dot and prime are derivatives with respect to $\tau$ and $\sigma$ and $G$ represent the components of the background metric. In order to study the dynamics of string, we use the metric of Eq.\ref{stringmetric}. 
In the static gauge, the corresponding Lagrangian density takes the following form,
\begin{eqnarray}\label{Ldensity}
\mathcal{L} = -\frac{1}{2\pi \alpha'} \frac{e^{2 A_s(z)} }{z^2}\sqrt{1-\frac{\dot{x}^2}{g(z)}+g(z) x'(z)^2}\,.
\end{eqnarray}
Then the equation of motion for x is,
\begin{eqnarray}\label{eom}
\partial_t\left( \frac{\dot{x}}{\sqrt{-\gamma'}}\right)-\frac{z^2\, g(z)}{e^{2 A_s(z)}}\, \partial_z\left( \frac{e^{2 A_s(z)}g(z) x'}{z^2\sqrt{-\gamma'}}\right)=0 \,.
\end{eqnarray}
The static string stretching from the boundary to the horizon, $x(t,z)=Constant$  is a trivial solution of this equation. For a string whose endpoint at the boundary moves with a constant velocity $v$, the following ansatz is chosen,
\begin{equation}\label{HQprofile}
X = \{t, v t + \xi (z), z\}\,,
\end{equation}
which leads to the following equation of motion,
\begin{equation}\label{pi}
\frac{e^{2 A_s(z)} g(z) \xi '(z)}{\sqrt{-\gamma } z^2}\,=\,const\,\equiv \pi _x\,,
\end{equation}
where $\pi_x$ is the worldsheet conserved quantity. One can obtain the equation for $\xi$ as follows,
\begin{equation}\label{eomxi}
 \xi' (z)=\pm \frac{\pi _x }{g(z)}\sqrt{\frac{g(z)-v^2}{\frac{e^{4 A_s(z)} g(z)}{z^4}-\pi _x^2}}\,.
\end{equation}
Requiring the real values for the function under the square root fixes the constant of the equation of motion as,
\begin{eqnarray}\label{zs}
&& g(z_s)-v^2=0\, , \nonumber \\
&& \frac{e^{4 A_s(z_s)} g(z_s)}{z_s^4}-\pi _x^2=0  \, \,\, 	\Rightarrow  \, \,\,  \pi _x=  \frac{e^{2 A_s(z_s)} v}{z_s^2}\,.
\end{eqnarray}
Finally, substituting the Eq.\ref{zs} into the Eq.\ref{eomxi}, one can solve the string solution,
\begin{equation}\label{eomxi2}
 \xi' (z)=-  \frac{e^{2 A_s(z_s)} v}{z_s^2}\frac{1}{g(z)}\sqrt{\frac{g(z)-g(z_s)}{\frac{e^{4 A_s(z)} g(z)}{z^4}-\frac{e^{4 A_s(z_s)} g(z_s)}{z_s^4}}}\, .
\end{equation}
The canonical momentum densities associated to the string are,
\begin{eqnarray}
    \left( \begin{array}{c}
  \label{densities}
    \pi^0_t \\ \pi^0_x 
    \end{array}
    \right)
   & =&
    \frac{T_0 L^4}{\sqrt{-\gamma}} \,\frac{e^{4 A_s(z)}}{z^4 g(z)}
    \left(
    \begin{array}{c}
    -g(z)\, \left(g(z) \xi '(z)^2+1\right) \\
   v   \end{array}
    \right) ,
     \\ \nonumber
     \\ 
    \left(
  \begin{array}{c}
   \label{flows}
    \pi^1_t \\ \pi^1_x 
    \end{array}
    \right)
  &  =&
    \frac{T_0 L^4}{\sqrt{-\gamma}} \,\frac{e^{4 A_s(z)}}{z^4}
    \left(
    \begin{array}{c}
   -v \,g(z)\,\xi '(z) \\
   g(z)\,\xi '(z)
    \end{array}
    \right).
\end{eqnarray}
Integrating the $\pi^0_{t}$ and $\pi^0_{x}$ along the string gives us the total energy and the total momentum in the direction of motion of the string, respectively. While, $\pi^1_{t}$ and $\pi^1_{x}$ are the energy and momentum flow down along the string. It is straightforward to show that $\pi^1_{x}$ is exactly the constant of the equations of motions, $\pi_{x}$ in Eq.\ref{zs}. Also, just similar to the case of $\mathcal{N}=4$ SYM plasma, $ \pi^1_t =-v\, \pi^1_x $. It means that if we pull the quark with the constant velocity, the fraction of energy flow at a given point along the string, $ \pi^1_t$ is constant. This is the energy dissipating into the surrounding medium by the quark. Therefore, the drag force is obtained as,
\begin{equation}
\label{DragF}
F_{drag}= -\pi^1_x =-\frac{1} {2\pi \alpha'}\frac{e^{2 A_s(z_s)} v}{z_s^2}.
\end{equation}
The drag force of the AdS Schwarzschild black hole background can be obtained analytically as \cite{Gubser:2006qh,Herzog:2006gh},
\begin{equation}\label{AdSDrag}
F_{drag}^{SYM} = -\frac{\pi\, T^2 \, \sqrt{\lambda}}{2}\frac{v}{\sqrt{1-v^2}}     \,, 
\end{equation}
while in this background, we first need to solve the Eq.\ref{zs} numerically to obtain the $z_s$, and then calculate the drag force using the Eq.\ref{DragF}.\\

\begin{figure}
\begin{subfigure}{.45\textwidth}
%%%%%%%%%%
  \centering
  \includegraphics[width=0.95\linewidth]{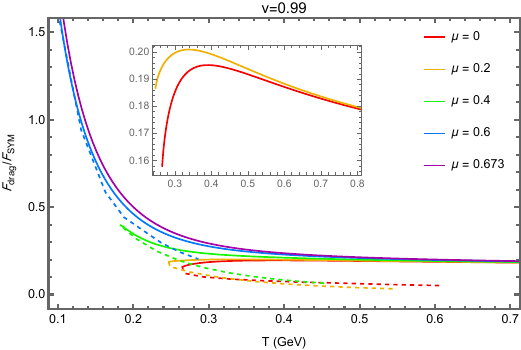}
   \caption{}
  \label{ratio1}
\end{subfigure}
%%%%%%%%%%
\begin{subfigure}{.45\textwidth}
  \centering
  \includegraphics[width=0.95\linewidth]{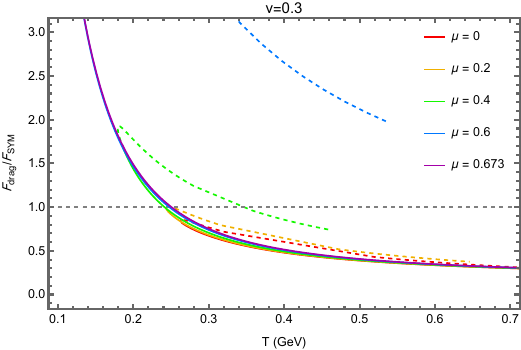}
   \caption{}
  \label{ratio2}
\end{subfigure}
%%%%%%%%%%
\caption{\small (Color online) The ratio of drag force in the holographic QCD model to its conformal value in terms of  temperature for $\mu=0, 0.2, 0.4, 0.6$ and $0.673$ (red, orange, green, blue and violet respectively). The solid curves correspond to the deconfined phase and the dashed curves represent the data of the confined phase. (\subref{ratio1}) The velocity of quark is $v=0.99$. (\subref{ratio2}) The velocity of quark is $v=0.3$. }
\label{DragForceRatio}
\end{figure}

\begin{figure}
\begin{subfigure}{.45\textwidth}
%%%%%%%%%%
  \centering
  \includegraphics[width=0.95\linewidth]{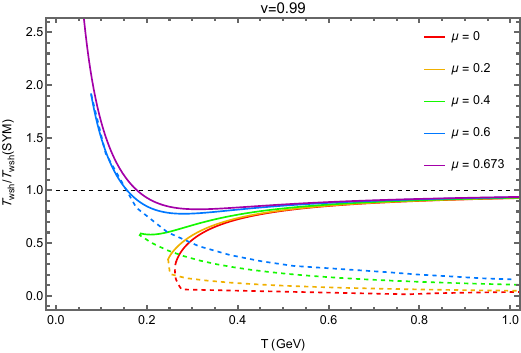}
   \caption{}
  \label{Twsh1}
\end{subfigure}
%%%%%%%%%%
\begin{subfigure}{.45\textwidth}
  \centering
  \includegraphics[width=0.95\linewidth]{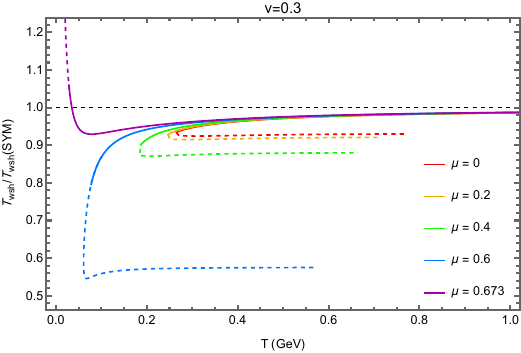}
   \caption{}
  \label{Twsh2}
\end{subfigure}
%%%%%%%%%%
\caption{\small (Color online) The ratio of the worldsheet temperature of the holographic QCD model to its conformal value in terms of temperature for $\mu=0, 0.2, 0.4, 0.6$ and $0.673$ (red, orange, green, blue and violet respectively). The solid curves correspond to the deconfined phase and the dashed curves represent the ratio in the confined phase. (\subref{ratio1}) The velocity of quark is $v = 0.99$. (\subref{ratio2}) The velocity of quark is $v = 0.3$. }
\label{Twsh}
\end{figure}

In \fig{DragForceRatio}, the ratio of drag force to its conformal value is plotted in terms of temperature for two different quark velocities and different values of $\mu$ from zero to the critical value of $0.673$. The solid curves represent the drag force ratio of the deconfined phase starting from the critical temperature $T_c$ while the non-physical dashed curves correspond to the confined phase drag ratio. As mentioned in section \ref{EMDG}, a first-order phase transition from thermal AdS to black hole phase occurs by decreasing $z_h$, and for each chemical potential value, the critical temperature is the temperature in which the free energy sign changes. Therefore, there exist two solutions at each temperature, one for the thermal AdS solution and the other for the AdS Black hole solution. The drag ratio develops a peak around the critical temperature only for the high velocity $v=0.99$. The peak moves towards higher temperatures for smaller values of chemical potential and higher quark velocities, as shown in red and orange curves inside Fig.\ref{ratio1}. For the deconfined phase, the drag force ratio increases by increasing the chemical potential value and decreasing the quark velocity. The same behavior in temperature and chemical potential has been reported in the holographic QCD model of \cite{Zhu:2021nbl} for the Einstein-Maxwell-scalar gravity system (also in \cite{Rougemont:2015wca} the same behavior on baryon chemical potential has been shown). From this figure, one could also find that at high temperatures, the curves of all chemical potential values converge and the drag force would be less than the value of $\mathcal{N}=4\,SYM$ theory due to the smaller ’t Hooft coupling of the holographic QCD model \footnote{In this paper, we have considered $T_s L^2=0.1$ \cite{Dudal:2017max} which corresponds to a small ’t Hooft coupling for the holographic QCD model compared to the $\mathcal{N}=4 SYM$ coupling that is considered to be $6\pi$.}. At lower temperatures, the drag force ratio is more sensitive to the chemical potential value, and the drag force of the holographic model is larger than the drag force of the $\mathcal{N}=4 SYM$ theory for larger $\mu$ values. For lower quark velocities, the drag force ratio is less sensitive to the chemical potential (Fig.\ref{ratio2}).\\
The world-sheet coordinates can be reparametrized as,
\begin{eqnarray}\label{rep_ws}
&& \tau\,=\,t+K(z) , \nonumber \\
&& x\,=\,v\,t+v\,K(z)+\xi(z) \, \, .
\end{eqnarray}
Since, the dynamic of the string is independent from $K(z)$, one can choose the following ansatz,
\begin{eqnarray}\label{K}
K'(z)\,=\,- \frac{G_{xx}\,v}{G_{tt}\,+\,G_{xx}\,v^2} \, \xi' (z) \, .
\end{eqnarray}
By substituting the background metric components and using Eqs.\ref{zs}, and \ref{eomxi2}, the induced metric simplifies as,

\begin{eqnarray}
h_{\alpha,\beta}\,=\,\frac{ L^2\,e^{2A_s(z)}  }{z^2    }
    \left( \begin{array}{cc}
  \label{diagonalinducedmetric}
   -(g(z)\,-v^2)                &     0    \\ 
            0                       &  \frac{z_s^4\,e^{4A_s(z)}}{ e^{4A_s(z)} \,g(z)\,z_s^4\,- \,e^{4A_s(z_s)\,g(z_s)\,z^4}  }
                \end{array}
    \right)
 \end{eqnarray}
which can be considered as the metric of a two-dimensional world-sheet black hole with the horizon radius of $z_s$. The local speed of light at the worldsheet horizon corresponds to the speed of a quark at the boundary. The associated Hawking temperature of the worldsheet is calculated numerically and its ratio to its conformal value is plotted in \fig{Twsh} for two different quark velocities. Similar to the drag force, the behavior of this ratio at lower temperatures, near the critical temperature, is very dependent on the chemical potential and quark velocity, such that for $\mu$ close to $\mu_c$ the worldsheet temperature in this background can be larger than its conformal value. However, at large temperatures, this ratio tends to one independent of the $\mu$ and quark velocity values. Similar to the drag force ratio plot of Fig.\ref{DragForceRatio}, for each chemical potential value, the solid curve starts from its critical temperature and the solid and dashed curves correspond to the deconfined and confined phases respectively.

%%%%%%%%%%%%%%%%%%%%%%%%%%%%%%%%%%%%%%%%%%%%%%%%%%%%%%%%%%%%%%
\section{Langevin equation
\label{Langevin}}
%%%%%%%%%%%%%%%%%%%%%%%%%%%%%%%%%%%%%%%%%%%%%%%%%%%%%%%%%%%%%%

Heavy quarks moving at a constant velocity in the plasma experience a Brownian motion which can be described by the effective equation of motion \cite{Gubser:2006qh}
\begin{equation}\label{brownian}
\frac{dp}{dt}\,=\,-\,\eta_D \,p \, +\, \xi(t),
\end{equation}
where $\eta_D$ is the drag coefficient, $p$ is the relativistic expression of the momentum of the quark, and $\xi$ is a random force, expressing the interaction of the medium with the heavy quark. This random force causes the momentum broadening of the quark which can be extracted by analyzing small fluctuations in the path of the Wilson line. This is dual to perturbing the location of the classical string endpoint on the boundary which yields fluctuations on the string world sheet dragging behind the quark. Therefore, we consider the quadratic fluctuations around the classical trailing string solution obtained in the previous section. In the static gauge, $\tau = t$, and $\sigma = z$, we generalise the embedding function of the string as 
\begin{equation}\label{generalizedembedding}
x_L(t,z)=v\,t\,+\xi(z)+\delta\, x(t,z) \,\,\,\, ,\,\,\, x_T= \delta\,x_T(t,z)     \,,
\end{equation}
where $L$ and $T$ denote to the parallel and transverse to the direction of motion. Now, we rewrite the Nambu-Goto action Eq.\ref{NGaction} and expand it to second order in terms of fluctuations around the string solution, Eq.\ref{eomxi2}
\begin{eqnarray}\label{NGaction2}
S= -\frac{1}{2\pi \alpha'} \int\,dz \,dt \,\sqrt{-\gamma} \frac{\gamma^{\alpha\beta}}{2}\, \left[  N(z)\,  \partial_{\alpha} \delta x_L  \,  \partial_{\beta} \delta x_L   +      G_{xx} \,  \partial_{\alpha} \delta x_T  \,  \partial_{\beta} \delta x_T    \right],
\end{eqnarray}
where 
\begin{eqnarray}\label{gamma}
\gamma = \frac{L^4 z_s^4}{z^4} \frac{e^{8A_s(z)} (g(z)-v^2)}{ e^{4A_s(z)} g(z)z_s^4 -  e^{4A_s(z_s)} g(z_s) z^4   } \, ,
\end{eqnarray}
and 
\begin{eqnarray}\label{n}
N(z)= \frac{L^2 }{e^{2As(z)}z^2} \frac{e^{4A_s(z)} g(z)  z_s^4 -  e^{4A_s(z_s)} g(z_s) z^4    }{ z_s^4  (g(z)-v^2)    } \, .
\end{eqnarray}

The above action can be rewritten in terms of the worldsheet coordinate Eq.\ref{rep_ws} and ansatz Eq.\ref{K} which diagonalized the induced metric, Eq.\ref{diagonalinducedmetric}
\begin{eqnarray}\label{NGaction3}
S= -\frac{1}{2\pi \alpha'} \int\,dz \,dt \,\frac{H^{\alpha\beta}}{2}\, \left[  N(z)\,  \partial_{\alpha} \delta x_L  \,  \partial_{\beta} \delta x_L   +      G_{TT} \,  \partial_{\alpha} \delta x_T  \,  \partial_{\beta} \delta x_T    \right],
\end{eqnarray}
where $H^{\alpha\beta}$ is defined in terms of inverse of the diagonalized induced metric Eq.\ref{diagonalinducedmetric} as
\begin{equation}\label{H}
H^{\alpha\beta} = \sqrt{h}   h^{\alpha\beta}   \,,
\end{equation}
One can read the transport coefficient directly from the above action according to the membrane paradigm \cite{Iqbal:2008by}. The quadratic effective action for a massless scalar field $\phi$ has the form of
\begin{eqnarray}\label{Sphi}
S= -\frac{1}{2} \int\,dz \,dt \,  \sqrt{-g}\,q(z)\, g^{MN} \partial_M \phi \,\partial_N \phi \,.
\end{eqnarray}
The momentum broadening coefficients can be read directly from the above action as 
\begin{eqnarray}\label{kappa}
\kappa= \lim\limits_{\omega \to 0} \left (-\frac{2 T_{wsh}}{\omega} \mathrm{Im} \hat{G}_R(\omega) \right) = 2 \,T_{wsh} \, q(z)   \,,
\end{eqnarray}
where  $\mathrm{Im} \hat{G}_R$ is the imaginary part of retarded correlation function at the world-sheet horizon.
By comparing the action \ref{NGaction3} with Eq.\ref{Sphi} and implying the relation \ref{kappa}, the Langevin coefficients can be read as \cite{CasalderreySolana:2011us,Gursoy:2010aa,Giataganas:2013hwa}
\begin{eqnarray}\label{kappa2}
\kappa_T&=& \frac{1}{\pi \alpha'}\,T_{wsh} \, \left. G_{xx} \right \vert_{z \rightarrow z_s}   \,\nonumber \\
\kappa_L&=& \frac{1}{\pi \alpha'}\,T_{wsh} \, \left. N(z) \right \vert_{z \rightarrow z_s} 
\end{eqnarray}
The final results for the ratio of the longitudinal to transverse transport coefficients in this model is 
\begin{equation}\label{kratio}
\frac{\kappa_L}{\kappa_T}=1+\frac{ 4g(z_s)\left(z_s A_s'(z_s)-1\right)} { z_s g'(z_s)  },
\end{equation}
where the L’ Hospital’s rule were employed since our functions are continuous and $z_s$ is calculated using Eq.\ref{zs}.\\

\begin{figure}
\begin{subfigure}{.45\textwidth}
%%%%%%%%%%
  \centering
  \includegraphics[width=0.95\linewidth]{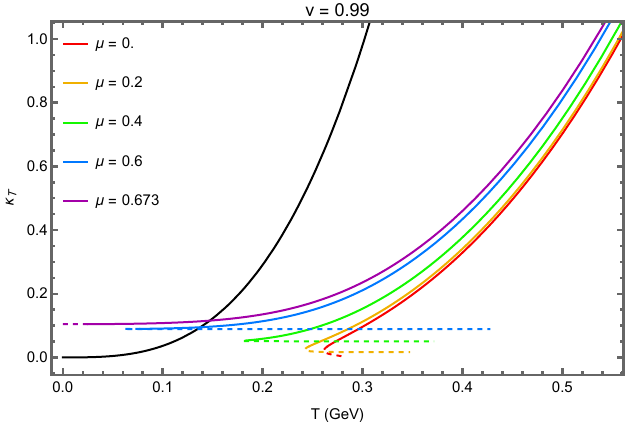}
   \caption{}
  \label{k1}
\end{subfigure}
%%%%%%%%%%
\begin{subfigure}{.45\textwidth}
  \centering
  \includegraphics[width=0.95\linewidth]{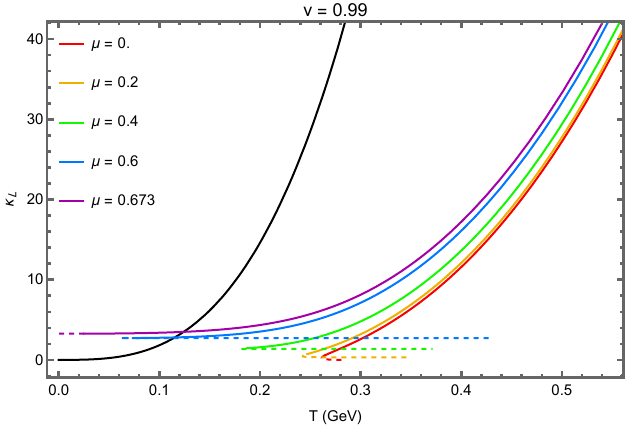}
   \caption{}
  \label{k2}
\end{subfigure}\\
%%%%%%%%%%
\begin{subfigure}{.45\textwidth}
%%%%%%%%%%
  \centering
  \includegraphics[width=0.95\linewidth]{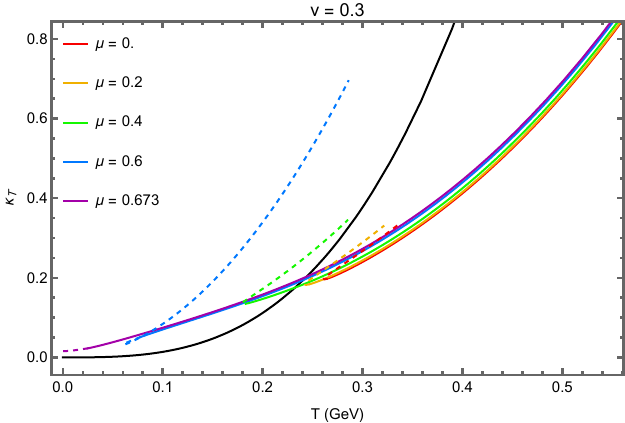}
   \caption{}
  \label{k3}
\end{subfigure}
%%%%%%%%%%
\begin{subfigure}{.45\textwidth}
  \centering
  \includegraphics[width=0.95\linewidth]{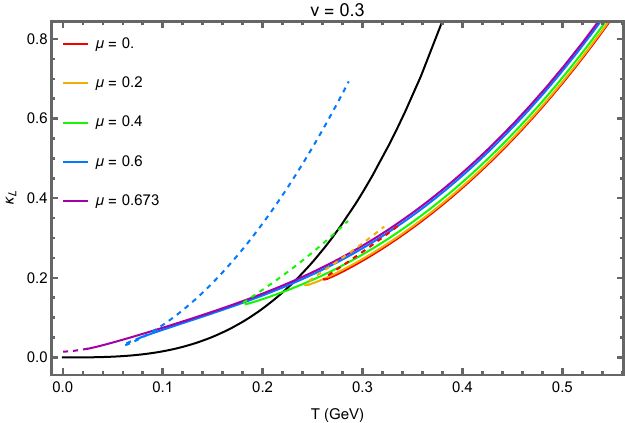}
   \caption{}
  \label{k4}
\end{subfigure}
\caption{\small (Color online) The longitudinal and transverse transport coefficients in the holographic model in terms of the temperature for $\mu=0, 0.2, 0.4, 0.6$ and $0.673$ (red, orange, green, blue and violet respectively). (\subref{k1}) $\kappa_T$ for $v=0.99$, (\subref{k2})  $\kappa_L$ for $v=0.99$, (\subref{k3}) $\kappa_T$ for $v=0.3$ and (\subref{k4}) $\kappa_L$ for $v=0.3$.
\label{kTL}}
\end{figure}

In Fig.\ref{kTL}, we have plotted the transverse and longitudinal transport coefficients versus temperature for two different quark velocities. In these plots, the black solid curves represent the $\kappa_T$ and $\kappa_L$ of the $\mathcal{N}=4\, SYM$ theory, and the dashed lines represent the data of the confined phase of the holographic model. From the deconfined curves, we find that increasing the temperature and chemical potential, increases the transport coefficients and for high quark velocity, the dependency on $\mu$ is more relevant. Also, one could observe that at lower temperatures (available for higher chemical potential values) the transverse and longitudinal transport coefficients are larger than the $\mathcal{N}=4\, SYM$ while for smaller $\mu$ values, they are smaller than $\mathcal{N}=4\, SYM$. Our numerical results for the transport coefficients obey the expected inequality in isotropic backgrounds $\kappa_L >\kappa_T$ \cite{Giataganas:2013hwa}.

\begin{figure}
\begin{subfigure}{.45\textwidth}
%%%%%%%%%%
  \centering
  \includegraphics[width=0.95\linewidth]{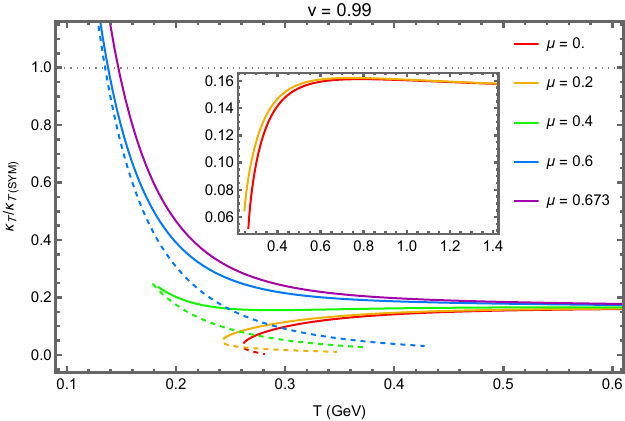}
   \caption{}
  \label{kratio1}
\end{subfigure}
%%%%%%%%%%
\begin{subfigure}{.45\textwidth}
  \centering
  \includegraphics[width=0.95\linewidth]{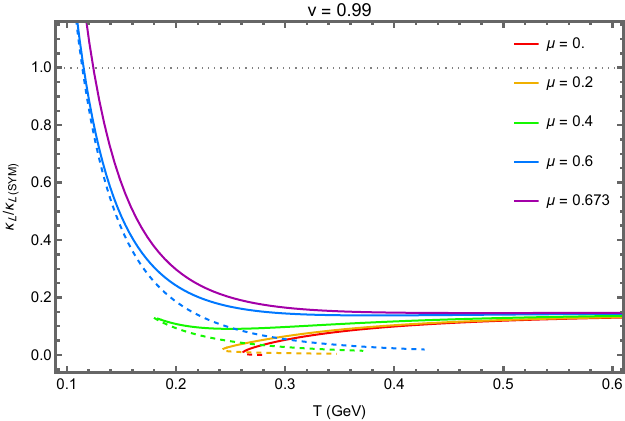}
   \caption{}
  \label{kratio2}
\end{subfigure}\\
%%%%%%%%%%
\begin{subfigure}{.45\textwidth}
%%%%%%%%%%
  \centering
  \includegraphics[width=0.95\linewidth]{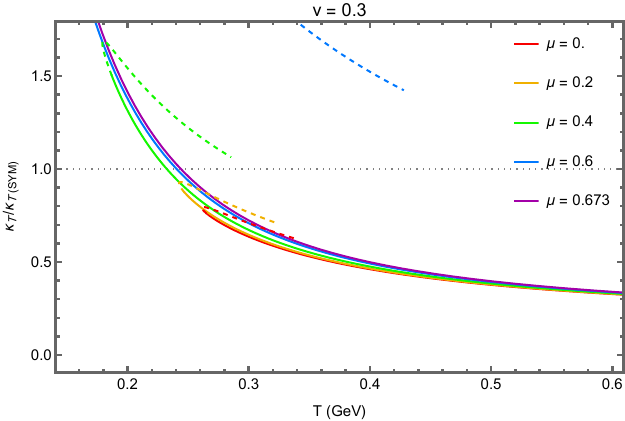}
   \caption{}
  \label{kratio3}
\end{subfigure}
%%%%%%%%%%
\begin{subfigure}{.45\textwidth}
  \centering
  \includegraphics[width=0.95\linewidth]{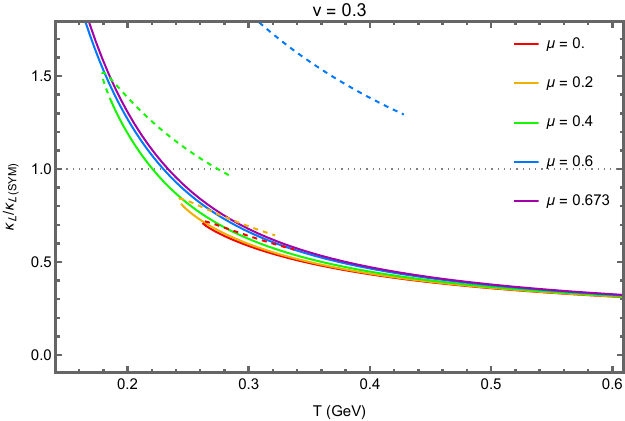}
   \caption{}
  \label{kratio4}
\end{subfigure}
\caption{\small (Color online) The ratio of the longitudinal to transverse transport coefficients in the holographic model in terms of the temperature for $\mu=0, 0.2, 0.4, 0.6$ and $0.673$ (red, orange, green, blue and violet respectively). (\subref{kratio1}) The transverse ratio for $v=0.99$, (\subref{kratio2})  The longitudinal for $v=0.99$, (\subref{kratio3}) The transverse ratio for $v=0.3$ and (\subref{kratio4}) the longitudinal ratio for $v=0.3$.
\label{kTLratio}}
\end{figure}

The ratios of $\kappa_L/\kappa_{L(SYM)}$ and $\kappa_T/\kappa_{T(SYM)}$ in terms of the temperature for two quark velocities are shown in Fig.\ref{kTLratio}. At lower temperatures, the dependency on $\mu$ is more evident while at high temperatures, all the curves converge to a single value smaller than one. The data of each $\mu$, exhibit a peak around the critical temperature that moves towards higher temperatures for smaller values of chemical potential (except for the small $\mu$ curves of Fig.\ref{kratio2} in which no peaks exist). From these plots, one could find that the transport coefficients increase by increasing the chemical potential similar to the transport coefficients in a  medium with baryon density \cite{Rougemont:2015wca}.

%%%%%%%%%%%%%%%%%%%%%%%%%%%%%%%%%%%%%%%%%%%%%%%%%%%%%%%%%%%%%%

\section{Jet Quenching Parameter
\label{JetQ}}

In this section, we study the light quark energy loss by calculating the jet quenching parameter. For this purpose, we follow the method of \cite{Liu:2006ug} to compute this parameter in the holographic QCD model of \cite{Dudal:2018ztm,Dudal:2017max} reviewed in section \ref{EMDG}.\\ 
Starting from the string frame metric of Eq.\ref{stringmetric} and using the light-cone coordinates $x^{\pm}=(x_1\pm t)/\sqrt{2}$, the bulk metric takes the following form,
 \begin{eqnarray}
ds^2= \frac{L^2 \, e^{2 A_s(z)}}{ z^2} \left( d{x_2}^ 2 + d{x_3}^2+\frac{1}{2}\left(1-g(z)\right) \left(d{x^+}^ 2 + d{x^-}^2 \right) +\left(1+g(z)\right)dx^+ dx^-  +\frac{ dz^2}{g(z)}\right) .
\label{eq:metric-l-c}
 \end{eqnarray}
To calculate the jet quenching parameter, the thermal expectation value of a closed rectangular Wilson loop is being used as \cite{Liu:2006ug},
\begin{equation}
\langle{W^A({\cal C})}\rangle  \approx \exp \left[ - \frac{1}{4\sqrt{2}}  \hat{q}\, L^-\, L^2 \right]\, ,
\label{eq:wilson1}
\end{equation}
where $L^-$ is the distance (conjugate to partons with relativistic velocities) and $L$ is the transverse distance (conjugate to the transverse momentum of the radiated gluons). This equation is valid for  $L^-\gg L$. On the other hand, Thermal expectation value of the Wilson loop $\langle{W^F({\cal C})}\rangle$ is obtained by using the extremal  surface action as \cite{Maldacena:1998im,Rey:1998ik,Rey:1998bq,Brandhuber:1998bs,Sonnenschein:1999if}      
\begin{equation}
\langle{W^F({\cal C})}\rangle = \exp \left[ -  S_I({\cal C})\right] \, ,
\label{eq:wilson2}
\end{equation} 
where $S_I$ is the normalized action of a hanging string from Wilson loop $\cal C$  joining two light-like lines. to the bulk (after subtracting the self energy of the $q\bar{q}$ pair from the Nambu-Goto action of the string worldsheet). In the large $N_c$ limit, one can use the relation ${\rm Tr}_{(Adj.)} = {\rm Tr^2}_{(Fund.)}$ and compare the equations \ref{eq:wilson1} and \ref{eq:wilson2} to obtain,  
\begin{equation}
\hat{q}=\frac{8\sqrt{2}S_I}{L^-\, L^2}.
\label{eq:q-SI}
\end{equation} 
By the string parametrization $x^{\mu}(\tau,\sigma)$, the Nambu-Goto action of the string is written as,
\begin{equation}
S_{NG} ={1 \over 2 \pi \alpha'} \int d\sigma d \tau \, \sqrt{ -\det \gamma_{\alpha\beta}},
\label{eq:NG}
 \end{equation}
where $\gamma_{\alpha \beta}$ is the induced metric on the string worldsheet and the worldsheet coordinates $\sigma^\alpha=(\tau,\sigma)$ are set to be $(x^-,x_2)$. The contour length along $x_2$-direction is defined by $L$ and the length along $\tau$-direction by $L^-$. The boundary conditions are $\phi(\pm\frac{L}{2})=0$ and $x_3(\sigma)$ and $x^+(\sigma)$ coordinates are constant. Then the action of Eq.\ref{eq:NG} reads,
\begin{equation}
 S_{NG} = \frac{L^-}{\sqrt{2} \pi \alpha'}  \int_0^{{L \over 2}} d{\sigma} \,
\frac{e^{2A_s(z)}}{z^2} \sqrt{\frac{1-g(z)}{2} \left(1+\frac{z'^2}{g(z)}\right)}\ ,
\label{eq:NG2}
\end{equation}
where prime denotes the derivative with respect to $x_2$. Since the Lagrangian density is time independent, the Hamiltonian of the system is constant,
\begin{equation}
\mathcal{L}-z' \frac{\partial \mathcal{L}}{\partial z'}=\frac{\Pi_{z}}{\sqrt{2}}.
\label{eq:H}
\end{equation}
From the above equation, one can obtain $z'$ as,
\begin{equation}
z' =\sqrt{g(z) \left( \frac{e^{4A_s(z)}(1-g(z))}{{\Pi_{z}}^2 z^4}-1 \right)} .
\label{eq:zp}
\end{equation} 
Integrating equation \ref{eq:zp} leads to,
\begin{equation}
\frac{L}{2} = a_0 {\Pi_{z}} +\mathcal{O}({\Pi_{z}}^3), 
\label{eq:L}
\end{equation}
where,
\begin{equation}
a_0=\int_{z_h}^{{0}} d z \frac{z^2 e^{-2A_s(z)}}{\sqrt{g(z) \left(1-g(z)\right)}}\,.
\label{eq:a0}
\end{equation}
Here, we have considered that for small length $L$, the constant $\Pi_{z}$ is small and its higher order terms are negligible. 
Substituting Eq.\ref{eq:zp} into Eq.\ref{eq:NG2} yields,
\begin{equation}
S_{NG}=\frac{L^-}{\pi \alpha'}\int_{z_H}^{{0}} d z \frac{ e^{4A_s(z)} (1-g(z))}{z^2 \sqrt{2g(z) (e^{4A_s(z)}(1-g(z))+\Pi_z^{2} z^4)}},
\label{eq:NG3}
\end{equation} 
where we have used $z' =\frac{\partial{z}}{\partial{\sigma}}$. Expanding this equation for small $\Pi_z$, leads to,
\begin{equation}
S_{NG}=\frac{L^-}{\pi \alpha'}\int_{z_H}^{{0}} d z \frac{ e^{2A_s(z)}}{z^2} \sqrt{\frac{1-g(z)}{2g(z)}}\left(1+\frac{e^{-4A_s(z)} \Pi_z^{2} z^4}{2(1-g(z))}+ ...\right),
\label{eq:NG4}
\end{equation} 
This action diverges and one should subtract the self energy of two disconnected strings whose worldsheets are located at $x_2=\pm\frac{L}{2}$ and extended from boundary to horizon as,
\begin{equation}
S_{0}=\frac{L^-}{2 \pi \alpha'}\int_{z_H}^{{0}} d z \sqrt{g_{--}g_{zz}}=\frac{L^-}{2 \pi \alpha'}\int_{z_H}^{{0}} d z \,\frac{e^{2 A_s(z)}}{z^2}\sqrt{\frac{1-g(z)}{2 g(z)}}\,.
\label{eq:S0}
\end{equation} 
The normalized action is therefore,
\begin{equation}
S_{I}=S_{NG}-2 S_0\equiv\frac{L^- \Pi_z^{2} a_0}{2\sqrt2\pi \alpha'}\,.
\label{eq:SI}
\end{equation} 
Inserting Eq.\ref{eq:SI} into Eq.\ref{eq:q-SI} leads to the following expression for the jet quenching parameter of holographic model,
\begin{equation}
\hat{q}=\frac{1}{\pi \alpha' a_0},
\label{eq:qfinal}
\end{equation} 
where we have used Eq.\ref{eq:L} for $\Pi_z$ and $a_0$ is the numerical integral defined in Eq.\ref{eq:a0}. For $\mathcal N=4$ supersymmetric Yang-Mills theory in the large $N_c$ and large $\lambda$ limits, Eq.\ref{eq:qfinal} leads to the following analytical equation \cite{Liu:2006ug},
\begin{equation}
\hat{q}_{SYM}=\frac{\pi^{3/2}\Gamma(\frac{3}{4})}{\Gamma(\frac{5}{4})}\sqrt{\lambda} \,T^3,
\label{eq:qSYM}
\end{equation} 

\begin{figure}
%%%%%%%%%%
\centering
\includegraphics[width=0.5\linewidth]{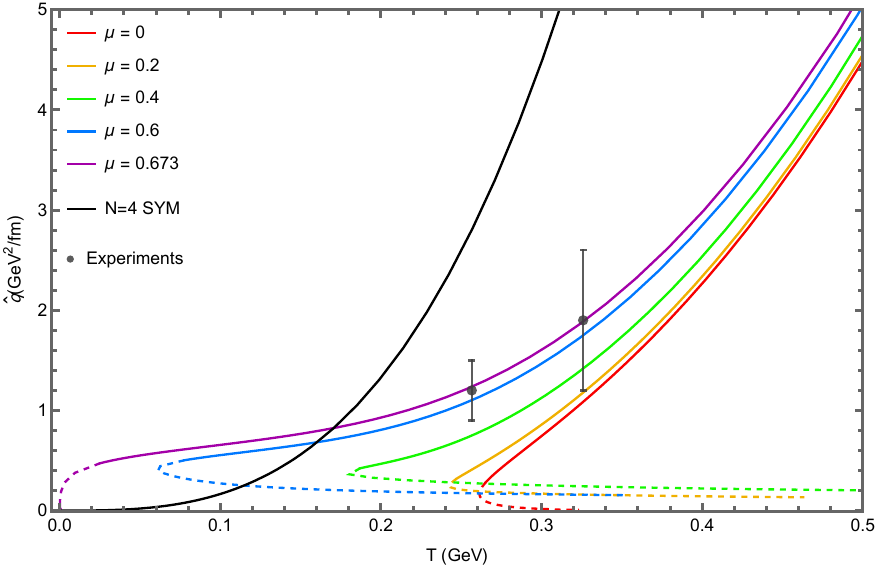}
\caption{\small (Color online) Jet quenching parameter versus temperature for the holographic model with different chemical potentials. The black curve is $\hat{q}_{SYM}$ and the circles indicate the experimental values from RHIC and LHC. The dashed curves represent the jet quenching parameter for the confined phase and the colored solid curves stand for the deconfined phase of the holographic model.
\label{qplot}}
\end{figure}

\begin{figure}
\begin{subfigure}{.45\textwidth}
%%%%%%%%%%
  \centering
  \includegraphics[width=0.95\linewidth]{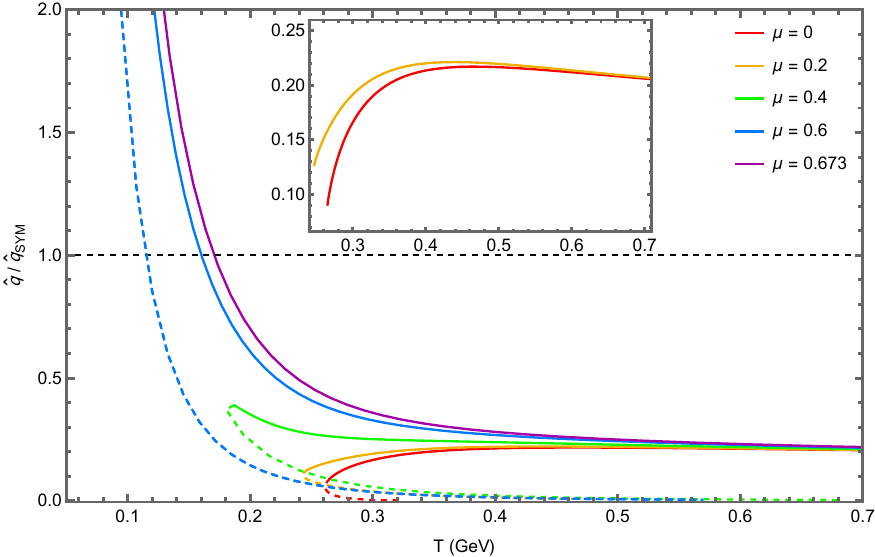}
   \caption{}
  \label{qrT}
\end{subfigure}
%%%%%%%%%%
\begin{subfigure}{.45\textwidth}
  \centering
  \includegraphics[width=0.95\linewidth]{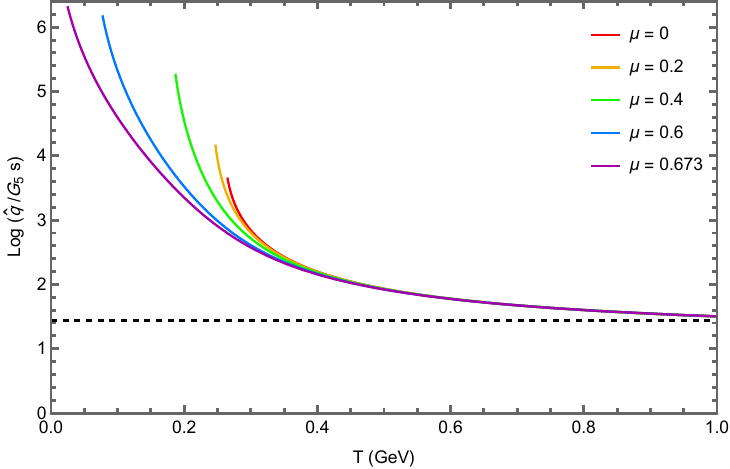}
   \caption{}
  \label{qsT}
\end{subfigure}
\caption{\small (Color online) (\subref{qrT}) $\hat{q}/\hat{q}_{SYM}$ and (\subref{qsT}) $Log(\hat{q}/G_5 s)$ versus temperature for the holographic QCD model. Red, orange, green, blue and violet curves correspond to $\mu=0, 0.2, 0.4, 0.6$ and $0.673$ respectively).
\label{qrs}}
\end{figure}

In order to obtain the jet quenching parameter for the holographic QCD model of \cite{Dudal:2017max}, we have solved the equation of \ref{eq:qfinal} numerically for different values of temperature ($z_h$) and chemical potential. The resulting curves are plotted in Fig.\ref{qplot}. In this figure, the numerical values of the jet quenching parameter are shown for the holographic QCD model (red, orange, green, blue, and purple curves with $\mu=0,\, 0.2,\, 0.4,\, 0.6$ and $\mu_c$ respectively). Solid curves correspond to the deconfined phase of the holographic model and dashed curves represent the confined phase data. For each chemical potential value, the solid curve starts from the critical temperature $T_c$, at which the confinement/deconfinement phase transition occurs. In this figure, the solid black curve displays the $\hat{q}_{SYM}$ and the black circles with error bars are the absolute values of $\hat{q}$ for a 10 GeV quark jet in the most central Au-Au collisions at RHIC with the highest temperature $T=0.37 GeV$ and Pb-Pb collisions at LHC with the highest temperature $T=0.47 GeV$ \cite{Burke:2013yra}\footnote{In this plot, these two temperatures are rescaled as $T\approx T_{SYM}=3^{-1/3} T_{QCD}$ since for the holographic QCD theories, the number of degrees of freedom is more than those of 3 favor QCD \cite{Burke:2013yra}.}. From this figure, one can observe that increasing the chemical potential and temperature leads to an increase of the jet quenching parameter. At lower temperatures associated with larger values of chemical potential, $\hat{q}>\hat{q}_{SYM}$ and at higher temperatures, $\hat{q}<\hat{q}_{SYM}$. Experimental values of the jet quenching parameter, are closer to the higher chemical potential curves.\\
The ratio of the jet quenching parameter in the holographic QCD model to the $\hat{q}_{SYM}$ is plotted in Fig.\ref{qrT} in terms of temperature for $\mu=0,\, 0.2,\, 0.4,\, 0.6$ and $\mu_c$. From this figure, one could observe that at higher temperatures, the jet quenching curves converge to a value lower than the corresponding value of the $\mathcal N=4$ SYM theory similar to the curves of the drag force and Langevine coefficients in Figs.\ref{DragForceRatio} and \ref{kTLratio}. However, at lower temperatures, the jet quenching parameter of the holographic QCD model is larger than the $\hat{q}_{SYM}$. Similar to Figs.\ref{ratio1} and \ref{k1} for a high-velocity heavy quark, increasing the chemical potential value, increases this ratio, and the $\hat{q}/\hat{q}_{SYM}$ plot indicates a smooth confinement/deconfinement phase transition that results in a peak around the critical temperature for $\mu<\mu_c$ (for smaller $\mu$'s, the peak shifts to higher temperatures). The same behavior has been observed for $\hat{q}/T^3$ in the dynamical holographic QCD model of \cite{Li:2014hja} suggesting the rapid changing in the system degrees of freedom during the phase transition \cite{Rougemont:2015wca,Li:2014hja}.\\
Finally, the logarithmic plot of the jet quenching parameter over the entropy density is plotted in terms of the temperature in \fig{qsT} (in units of $G_5$) for different values of chemical potential. The dashed line represents the logarithmic value of $\hat{q}/G_5 s$ for the $\mathcal N=4$ SYM theory and each curve has a sharp rise with a finite peak at the crossover temperature. The peak increases by increasing the chemical potential and is higher than the $\hat{q}/ s$ curves of \cite{Li:2014hja}. It is worth mentioning that, despite the different concept and calculation methods, the jet quenching parameter of a light quark has the same behavior as the transverse transport coefficient of a high-velocity heavy quark (and consequently, to the Langevin jet quenching parameter $\hat{q}_T=2\kappa_T/v$).\\

\section{Conclusion
\label{conclusion}}
%%%%%%%%%%%%%%%%%%%%%%%%%%%%%%%%%%%%%%%%%%%%%%%%%%%%%%%%%%%%%%

In this paper, we studied the drag force and the Langevin diffusion coefficients for a heavy quark and also, the jet quenching parameter for a light quark moving through the QGP. For this purpose, we have implemented the holographic QCD model of \cite{Dudal:2017max} as a realistic framework for the QCD confinement/deconfinement phase transition with thermodynamic properties consistent with QCD and lattice data.\\
Our results demonstrate a nontrivial behavior on temperature and chemical potential, especially near the phase transition temperature. For a high-velocity heavy quark, the drag force and the transverse transport coefficient ratios increase by increasing the chemical potential. For each chemical potential value, these ratios develop a peak around the critical temperature that shifts to higher temperatures for higher quark velocities and smaller values of chemical potential. The longitudinal transport coefficients exhibit the same behavior except for the lower values of the chemical potential where no peaks exist. For a low-velocity quark, the drag force and transport coefficients ratios, monotonically decrease by increasing temperature and are less sensitive to the chemical potential values.\\
Also, we have calculated the jet quenching parameter for a light-like trajectory in terms of temperature for different values of chemical potential. This parameter, increases by increasing the chemical potential and temperature. Our results are in good agreement with the experimental data from RHIC and LHC. The ratios of $\hat{q}/\hat{q}_{SYM}$ are more than one at lower temperatures (accessible to the higher values of chemical potential). Similar to the drag force and transport coefficients, the curves develop peaks around the crossover temperature and at high $T$'s, converge to a single value less than one due to the small ’t Hooft coupling of the holographic QCD model we have implemented here. The jet quenching parameter of a light quark and the transverse transport coefficient of a high-velocity heavy quark, display the same qualitative behavior despite the differences. In the end, the ratio of $\hat{q}/G_5 s$ has been computed in terms of temperature for different values of chemical potential. The curves have sharp rising by decreasing the temperature and develop finite peaks at the crossover temperature, indicating that the system degrees of freedom change rapidly during the phase transition. 

%%%%%%%%%%%%%%%%%%%%%%%%%%%%%%%%%%%%%%%%%%%%%%%%%%%%%%
%%%%%%%%%%%%%%%%%%%%%%%%%%%%%%%%%%%%%%%%%%%%%%%%%%%%%%

\section*{Acknowledgement}
The authors wish to thank Hesam Soltanpanahi for the useful discussions and comments.

%%%%%%%%%%%%%%%%%%%%%%%%%%%%%%%%%%%%%%%%%%%%%%%%%%%%%%%%%%%%%%
\providecommand{\href}[2]{#2}\begingroup\raggedright
{}


\begin{thebibliography}{10}
%%%%%%%%%%%%%%%%%%%%%%%%%%%%%%%%%%%%%%%%%%%%%%%%%%%%%%%%%%%%%%

%\cite{Baier:1996kr}
\bibitem{Baier:1996kr} 
  R.~Baier, Y.~L.~Dokshitzer, A.~H.~Mueller, S.~Peigne and D.~Schiff,
  %``Radiative energy loss of high-energy quarks and gluons in a finite volume quark - gluon plasma,''
  Nucl.\ Phys.\ B {\bf 483}, 291 (1997)
  doi:10.1016/S0550-3213(96)00553-6
  [hep-ph/9607355].
  %%CITATION = doi:10.1016/S0550-3213(96)00553-6;%%
  %836 citations counted in INSPIRE as of 28 Dec 2018


%\cite{Eskola:2004cr}
\bibitem{Eskola:2004cr} 
  K.~J.~Eskola, H.~Honkanen, C.~A.~Salgado and U.~A.~Wiedemann,
  %``The Fragility of high-p(T) hadron spectra as a hard probe,''
  Nucl.\ Phys.\ A {\bf 747}, 511 (2005)
  doi:10.1016/j.nuclphysa.2004.09.070
  [hep-ph/0406319].
  %%CITATION = doi:10.1016/j.nuclphysa.2004.09.070;%%
  %265 citations counted in INSPIRE as of 28 Dec 2018

%\cite{Maldacena:1997re}
\bibitem{Maldacena:1997re} 
  J.~M.~Maldacena,
  %``The Large N limit of superconformal field theories and supergravity,''
  Int.\ J.\ Theor.\ Phys.\  {\bf 38}, 1113 (1999)
  [Adv.\ Theor.\ Math.\ Phys.\  {\bf 2}, 231 (1998)]
  doi:10.1023/A:1026654312961, 10.4310/ATMP.1998.v2.n2.a1
  [hep-th/9711200].
  %%CITATION = doi:10.1023/A:1026654312961, 10.4310/ATMP.1998.v2.n2.a1;%%
  %14269 citations counted in INSPIRE as of 28 Dec 2018


%\cite{Witten:1998qj}
\bibitem{Witten:1998qj} 
  E.~Witten,
  %``Anti-de Sitter space and holography,''
  Adv.\ Theor.\ Math.\ Phys.\  {\bf 2}, 253 (1998)
  doi:10.4310/ATMP.1998.v2.n2.a2
  [hep-th/9802150].
  %%CITATION = doi:10.4310/ATMP.1998.v2.n2.a2;%%
  %9276 citations counted in INSPIRE as of 28 Dec 2018


%\cite{Gubser:1998bc}
\bibitem{Gubser:1998bc} 
  S.~S.~Gubser, I.~R.~Klebanov and A.~M.~Polyakov,
  %``Gauge theory correlators from noncritical string theory,''
  Phys.\ Lett.\ B {\bf 428}, 105 (1998)
  doi:10.1016/S0370-2693(98)00377-3
  [hep-th/9802109].
  %%CITATION = doi:10.1016/S0370-2693(98)00377-3;%%
  %7934 citations counted in INSPIRE as of 28 Dec 2018


%\cite{Aharony:1999ti}
\bibitem{Aharony:1999ti} 
  O.~Aharony, S.~S.~Gubser, J.~M.~Maldacena, H.~Ooguri and Y.~Oz,
  %``Large N field theories, string theory and gravity,''
  Phys.\ Rept.\  {\bf 323}, 183 (2000)
  doi:10.1016/S0370-1573(99)00083-6
  [hep-th/9905111].
  %%CITATION = doi:10.1016/S0370-1573(99)00083-6;%%
  %4425 citations counted in INSPIRE as of 28 Dec 2018

%\cite{CasalderreySolana:2011us}
\bibitem{CasalderreySolana:2011us} 
  J.~Casalderrey-Solana, H.~Liu, D.~Mateos, K.~Rajagopal and U.~A.~Wiedemann,
  %``Gauge/String Duality, Hot QCD and Heavy Ion Collisions,''
  book:Gauge/String Duality, Hot QCD and Heavy Ion Collisions. Cambridge, UK: Cambridge University Press, 2014
  doi:10.1017/CBO9781139136747
  [arXiv:1101.0618 [hep-th]].
  %%CITATION = doi:10.1017/CBO9781139136747;%%
  %552 citations counted in INSPIRE as of 28 Dec 2018

%\cite{Polchinski:2000uf}
\bibitem{Polchinski:2000uf} 
  J.~Polchinski and M.~J.~Strassler,
  %``The String dual of a confining four-dimensional gauge theory,''
  hep-th/0003136.
  %%CITATION = HEP-TH/0003136;%%
  %613 citations counted in INSPIRE as of 28 Dec 2018
 
%\cite{Karch:2002sh}
\bibitem{Karch:2002sh} 
  A.~Karch and E.~Katz,
  %``Adding flavor to AdS / CFT,''
  JHEP {\bf 0206}, 043 (2002)
  doi:10.1088/1126-6708/2002/06/043
  [hep-th/0205236].
  %%CITATION = doi:10.1088/1126-6708/2002/06/043;%%
  %913 citations counted in INSPIRE as of 28 Dec 2018


%\cite{Sakai:2004cn}
\bibitem{Sakai:2004cn} 
  T.~Sakai and S.~Sugimoto,
  %``Low energy hadron physics in holographic QCD,''
  Prog.\ Theor.\ Phys.\  {\bf 113}, 843 (2005)
  doi:10.1143/PTP.113.843
  [hep-th/0412141].
  %%CITATION = doi:10.1143/PTP.113.843;%%
  %1185 citations counted in INSPIRE as of 28 Dec 2018


%\cite{He:2013qq}
\bibitem{He:2013qq}
S.~He, S.~Y.~Wu, Y.~Yang and P.~H.~Yuan,
%``Phase Structure in a Dynamical Soft-Wall Holographic QCD Model,''
JHEP \textbf{04}, 093 (2013)
doi:10.1007/JHEP04(2013)093
[arXiv:1301.0385 [hep-th]].
%93 citations counted in INSPIRE as of 17 Aug 2023

%\cite{Yang:2015aia}
\bibitem{Yang:2015aia}
Y.~Yang and P.~H.~Yuan,
%``Confinement-deconfinement phase transition for heavy quarks in a soft wall holographic QCD model,''
JHEP \textbf{12}, 161 (2015)
doi:10.1007/JHEP12(2015)161
[arXiv:1506.05930 [hep-th]].
%64 citations counted in INSPIRE as of 30 Aug 2023

\bibitem{Witten9803}
  E.~Witten,
  ``Anti-de Sitter space, thermal phase transition, and confinement in gauge theories,''
  Adv.\ Theor.\ Math.\ Phys.\  {\bf 2} (1998) 505
  [hep-th/9803131].

 \bibitem{Polchinski0003}
  J.~Polchinski and M.~J.~Strassler,
  ``The String dual of a confining four-dimensional gauge theory,''
  hep-th/0003136.

\bibitem{Sakai0412}
 T.~Sakai and S.~Sugimoto,
  ``Low energy hadron physics in holographic QCD,''
  Prog.\ Theor.\ Phys.\  {\bf 113} (2005) 843
  %doi:10.1143/PTP.113.843
  [hep-th/0412141].

\bibitem{Sakai0507}
 T.~Sakai and S.~Sugimoto,
  ``More on a holographic dual of QCD,''
  Prog.\ Theor.\ Phys.\  {\bf 114} (2005) 1083
  %doi:10.1143/PTP.114.1083
  [hep-th/0507073].


\bibitem{Kruczenski0311}
  M.~Kruczenski, D.~Mateos, R.~C.~Myers and D.~J.~Winters,
  ``Towards a holographic dual of large $N_c$ QCD,''
  JHEP {\bf 0405} (2004) 041
  %doi:10.1088/1126-6708/2004/05/041
  [hep-th/0311270].

\bibitem{Karch0205}
A.~Karch and E.~Katz,
  ``Adding flavor to AdS/CFT,''
  JHEP {\bf 0206} (2002) 043
  %doi:10.1088/1126-6708/2002/06/043
  [hep-th/0205236].

\bibitem{Klebanov0007}
  I.~R.~Klebanov and M.~J.~Strassler,
  ``Supergravity and a confining gauge theory: Duality cascades and chi SB resolution of naked singularities,''
  JHEP {\bf 0008} (2000) 052
  %doi:10.1088/1126-6708/2000/08/052
  [hep-th/0007191].

\bibitem{Erlich}
  J.~Erlich, E.~Katz, D.~T.~Son and M.~A.~Stephanov,
  ``QCD and a holographic model of hadrons,''
  Phys.\ Rev.\ Lett.\  {\bf 95} (2005) 261602
  %doi:10.1103/PhysRevLett.95.261602
  [hep-ph/0501128].

\bibitem{GursoyI}
 U.~Gursoy and E.~Kiritsis,
  ``Exploring improved holographic theories for QCD: Part I,''
  JHEP {\bf 0802} (2008) 032
  %doi:10.1088/1126-6708/2008/02/032
  [arXiv:0707.1324 [hep-th]].

\bibitem{GursoyII}
 U.~Gursoy, E.~Kiritsis and F.~Nitti,
  ``Exploring improved holographic theories for QCD: Part II,''
  JHEP {\bf 0802} (2008) 019
  %doi:10.1088/1126-6708/2008/02/019
  [arXiv:0707.1349 [hep-th]].

\bibitem{Gursoy}
 U.~Gursoy, E.~Kiritsis, L.~Mazzanti and F.~Nitti,
  ``Deconfinement and Gluon Plasma Dynamics in Improved Holographic QCD,''
  Phys.\ Rev.\ Lett.\  {\bf 101} (2008) 181601
  %doi:10.1103/PhysRevLett.101.181601
  [arXiv:0804.0899 [hep-th]].

  \bibitem{Gursoy:2010fj}
  U.~Gursoy, E.~Kiritsis, L.~Mazzanti, G.~Michalogiorgakis and F.~Nitti,
  ``Improved Holographic QCD,''
  Lect.\ Notes Phys.\  {\bf 828}, 79 (2011)
  %doi:10.1007/978-3-642-04864-7_4
  [arXiv:1006.5461 [hep-th]].

%\cite{Buchel:2006bv}
\bibitem{Buchel:2006bv} 
  A.~Buchel,
  %``On jet quenching parameters in strongly coupled non-conformal gauge theories,''
  Phys.\ Rev.\ D {\bf 74}, 046006 (2006)
  doi:10.1103/PhysRevD.74.046006
  [hep-th/0605178].
  %%CITATION = doi:10.1103/PhysRevD.74.046006;%%
  %93 citations counted in INSPIRE as of 28 Dec 2018

  \bibitem{Gursoy:2009jd}
  U.~Gursoy, E.~Kiritsis, L.~Mazzanti and F.~Nitti,
  ``Improved Holographic Yang-Mills at Finite Temperature: Comparison with Data,''
  Nucl.\ Phys.\ B {\bf 820}, 148 (2009)
 % doi:10.1016/j.nuclphysb.2009.05.017
  [arXiv:0903.2859 [hep-th]].


  \bibitem{Alho:2012mh}
  T.~Alho, M.~J\"arvinen, K.~Kajantie, E.~Kiritsis and K.~Tuominen,
  ``On finite-temperature holographic QCD in the Veneziano limit,''
  JHEP {\bf 1301}, 093 (2013)
  %doi:10.1007/JHEP01(2013)093
  [arXiv:1210.4516 [hep-ph]].

  \bibitem{Alho:2013hsa}
  T.~Alho, M.~J\"arvinen, K.~Kajantie, E.~Kiritsis, C.~Rosen and K.~Tuominen,
  ``A holographic model for QCD in the Veneziano limit at finite temperature and density,''
  JHEP {\bf 1404}, 124 (2014)
  Erratum: [JHEP {\bf 1502}, 033 (2015)]
 % doi:10.1007/JHEP02(2015)033, 10.1007/JHEP04(2014)124
  [arXiv:1312.5199 [hep-ph]].

  \bibitem{Jarvinen:2015ofa}
  M.~J\"arvinen,
  ``Massive holographic QCD in the Veneziano limit,''
  JHEP {\bf 1507}, 033 (2015)
  %doi:10.1007/JHEP07(2015)033
  [arXiv:1501.07272 [hep-ph]].

\bibitem{Herzog0608}
  C.~P.~Herzog,
  ``A Holographic Prediction of the Deconfinement Temperature,''
  Phys.\ Rev.\ Lett.\  {\bf 98} (2007) 091601
%  doi:10.1103/PhysRevLett.98.091601
  [hep-th/0608151].


\bibitem{Karch0602}
  A.~Karch, E.~Katz, D.~T.~Son and M.~A.~Stephanov,
  ``Linear confinement and AdS/QCD,''
  Phys.\ Rev.\ D {\bf 74} (2006) 015005
  %doi:10.1103/PhysRevD.74.015005
  [hep-ph/0602229].

\bibitem{Karch1012}
 A.~Karch, E.~Katz, D.~T.~Son and M.~A.~Stephanov,
  ``On the sign of the dilaton in the soft wall models,''
  JHEP {\bf 1104} (2011) 066
  %doi:10.1007/JHEP04(2011)066
  [arXiv:1012.4813 [hep-ph]].

\bibitem{Colangelo:2011sr}
  P.~Colangelo, F.~Giannuzzi, S.~Nicotri and V.~Tangorra,
  ``Temperature and quark density effects on the chiral condensate: An AdS/QCD study,''
  Eur.\ Phys.\ J.\ C {\bf 72} (2012) 2096
  %doi:10.1140/epjc/s10052-012-2096-9
  [arXiv:1112.4402 [hep-ph]].

\bibitem{Dudal:2015wfn}
  D.~Dudal, D.~R.~Granado and T.~G.~Mertens,
  ``No inverse magnetic catalysis in the QCD hard and soft wall models,''
  Phys.\ Rev.\ D {\bf 93} (2016) no.12,  125004
 % doi:10.1103/PhysRevD.93.125004
  [arXiv:1511.04042 [hep-th]].

  \bibitem{Dudal:2014jfa}
  D.~Dudal and T.~G.~Mertens,
  ``Melting of charmonium in a magnetic field from an effective AdS/QCD model,''
  Phys.\ Rev.\ D {\bf 91}, 086002 (2015)
  %doi:10.1103/PhysRevD.91.086002
  [arXiv:1410.3297 [hep-th]].

%\cite{Dudal:2015kza}
\bibitem{Dudal:2015kza}
  D.~Dudal and T.~G.~Mertens,
  ``Radiation Gauge in AdS/QCD: Inadmissibility and Implications on Spectral Functions in the Deconfined Phase,''
  Phys.\ Lett.\ B {\bf 751} (2015) 352.
  %doi:10.1016/j.physletb.2015.10.074
  [arXiv:1510.05490 [hep-th]].

  \bibitem{Callebaut:2011ab}
  N.~Callebaut, D.~Dudal and H.~Verschelde,
  ``Holographic rho mesons in an external magnetic field,''
  JHEP {\bf 1303}, 033 (2013)
  %doi:10.1007/JHEP03(2013)033
  [arXiv:1105.2217 [hep-th]].

\bibitem{Gherghetta:2009ac}
  T.~Gherghetta, J.~I.~Kapusta and T.~M.~Kelley,
  ``Chiral symmetry breaking in the soft-wall AdS/QCD model,''
  Phys.\ Rev.\ D {\bf 79} (2009) 076003
  [arXiv:0902.1998 [hep-ph]].


 \bibitem{Panero:2009tv}
  M.~Panero,
  ``Thermodynamics of the QCD plasma and the large-N limit,''
  Phys.\ Rev.\ Lett.\  {\bf 103}, 232001 (2009)
 % doi:10.1103/PhysRevLett.103.232001
  [arXiv:0907.3719 [hep-lat]].



\bibitem{Adare}
A.~Adare {\it et al.} [PHENIX Collaboration],
  ``$J/\psi$ Production vs Centrality, Transverse Momentum, and Rapidity in Au+Au Collisions at $\sqrt{s_{NN}} = 200$ GeV,''
  Phys.\ Rev.\ Lett.\  {\bf 98} (2007) 232301
  %doi:10.1103/PhysRevLett.98.232301
  [nucl-ex/0611020].

\bibitem{Abelev:2013ila}
  B.~B.~Abelev {\it et al.} [ALICE Collaboration],
  ``Centrality, rapidity and transverse momentum dependence of $J/\psi$ suppression in Pb-Pb collisions at $\sqrt{s_{\rm NN}}$=2.76 TeV,''
  Phys.\ Lett.\ B {\bf 734}, 314 (2014)
  %doi:10.1016/j.physletb.2014.05.064
  [arXiv:1311.0214 [nucl-ex]].


\bibitem{Matsui:1986dk}
  T.~Matsui and H.~Satz,
  ``$J/\psi$ Suppression by Quark-Gluon Plasma Formation,''
  Phys.\ Lett.\ B {\bf 178}, 416 (1986).


\bibitem{Kaczmarek:2005ui}
  O.~Kaczmarek and F.~Zantow,
  ``Static quark anti-quark interactions in zero and finite temperature QCD. I. Heavy quark free energies, running coupling and quarkonium binding,''
  Phys.\ Rev.\ D {\bf 71}, 114510 (2005)
  %doi:10.1103/PhysRevD.71.114510
  [hep-lat/0503017].

\bibitem{Hashimoto:2014fha}
  K.~Hashimoto and D.~E.~Kharzeev,
  ``Entropic destruction of heavy quarkonium in non-Abelian plasma from holography,''
  Phys.\ Rev.\ D {\bf 90}, no. 12, 125012 (2014)
  %doi:10.1103/PhysRevD.90.125012
  [arXiv:1411.0618 [hep-th]].

%\cite{Maldacena:1998im}
\bibitem{Maldacena:1998im} 
  J.~M.~Maldacena,
  %``Wilson loops in large N field theories,''
  Phys.\ Rev.\ Lett.\  {\bf 80}, 4859 (1998)
  doi:10.1103/PhysRevLett.80.4859
  [hep-th/9803002].
  %%CITATION = doi:10.1103/PhysRevLett.80.4859;%%
  %1561 citations counted in INSPIRE as of 28 Dec 2018


%\cite{Rey:1998ik}
\bibitem{Rey:1998ik} 
  S.~J.~Rey and J.~T.~Yee,
  %``Macroscopic strings as heavy quarks in large N gauge theory and anti-de Sitter supergravity,''
  Eur.\ Phys.\ J.\ C {\bf 22}, 379 (2001)
  doi:10.1007/s100520100799
  [hep-th/9803001].
  %%CITATION = doi:10.1007/s100520100799;%%
  %1223 citations counted in INSPIRE as of 28 Dec 2018


%\cite{Rey:1998bq}
\bibitem{Rey:1998bq} 
  S.~J.~Rey, S.~Theisen and J.~T.~Yee,
  %``Wilson-Polyakov loop at finite temperature in large N gauge theory and anti-de Sitter supergravity,''
  Nucl.\ Phys.\ B {\bf 527}, 171 (1998)
  doi:10.1016/S0550-3213(98)00471-4
  [hep-th/9803135].
  %%CITATION = doi:10.1016/S0550-3213(98)00471-4;%%
  %407 citations counted in INSPIRE as of 28 Dec 2018


 \bibitem{Iatrakis:2015sua}
  I.~Iatrakis and D.~E.~Kharzeev,
  ``Holographic entropy and real-time dynamics of quarkonium dissociation in non-Abelian plasma,''
  Phys.\ Rev.\ D {\bf 93}, no. 8, 086009 (2016)
 % doi:10.1103/PhysRevD.93.086009
  [arXiv:1509.08286 [hep-ph]].


  \bibitem{Fadafan:2015ynz}
  K.~Bitaghsir Fadafan and S.~K.~Tabatabaei,
  ``Entropic destruction of a moving heavy quarkonium,''
  Phys.\ Rev.\ D {\bf 94}, no. 2, 026007 (2016)
 % doi:10.1103/PhysRevD.94.026007
  [arXiv:1512.08254 [hep-ph]].

  \bibitem{Zhang:2016fwr}
  Z.~q.~Zhang, C.~Ma, D.~f.~Hou and G.~Chen,
  ``Entropic destruction of a rotating heavy quarkonium,''
  arXiv:1611.08011 [hep-th].

%\cite{Dudal:2018ztm}
\bibitem{Dudal:2018ztm} 
  D.~Dudal and S.~Mahapatra,
  %``Interplay between the holographic QCD phase diagram and entanglement entropy,''
  JHEP {\bf 1807}, 120 (2018)
  doi:10.1007/JHEP07(2018)120
  [arXiv:1805.02938 [hep-th]].
  %%CITATION = doi:10.1007/JHEP07(2018)120;%%
  %10 citations counted in INSPIRE as of 05 Aug 2019


%\cite{Dudal:2017max}
\bibitem{Dudal:2017max} 
  D.~Dudal and S.~Mahapatra,
  %``Thermal entropy of a quark-antiquark pair above and below deconfinement from a dynamical holographic QCD model,''
  Phys.\ Rev.\ D {\bf 96}, no. 12, 126010 (2017)
  doi:10.1103/PhysRevD.96.126010
  [arXiv:1708.06995 [hep-th]].
  %%CITATION = doi:10.1103/PhysRevD.96.126010;%%
  %7 citations counted in INSPIRE as of 20 Jun 2019

%\cite{Bohra:2019ebj}
\bibitem{Bohra:2019ebj}
H.~Bohra, D.~Dudal, A.~Hajilou and S.~Mahapatra,
%``Anisotropic string tensions and inversely magnetic catalyzed deconfinement from a dynamical AdS/QCD model,''
Phys. Lett. B \textbf{801}, 135184 (2020)
doi:10.1016/j.physletb.2019.135184
[arXiv:1907.01852 [hep-th]].
%61 citations counted in INSPIRE as of 08 Dec 2023


%\cite{Jena:2022nzw}
\bibitem{Jena:2022nzw}
S.~S.~Jena, B.~Shukla, D.~Dudal and S.~Mahapatra,
%``Entropic force and real-time dynamics of holographic quarkonium in a magnetic field,''
Phys. Rev. D \textbf{105}, no.8, 086011 (2022)
doi:10.1103/PhysRevD.105.086011
[arXiv:2202.01486 [hep-th]].
%6 citations counted in INSPIRE as of 08 Dec 2023


%\cite{Sommer:1993ce}
\bibitem{Sommer:1993ce}
R.~Sommer,
%``A New way to set the energy scale in lattice gauge theories and its applications to the static force and $\alpha_s$ in SU(2) Yang-Mills theory,''
Nucl. Phys. B \textbf{411}, 839-854 (1994)
doi:10.1016/0550-3213(94)90473-1
[arXiv:hep-lat/9310022 [hep-lat]].
%1003 citations counted in INSPIRE as of 27 Aug 2023}

%\cite{Adams:2005dq}
\bibitem{Adams:2005dq} 
  J.~Adams {\it et al.} [STAR Collaboration],
  %``Experimental and theoretical challenges in the search for the quark gluon plasma: The STAR Collaboration's critical assessment of the evidence from RHIC collisions,''
  Nucl.\ Phys.\ A {\bf 757}, 102 (2005)
  doi:10.1016/j.nuclphysa.2005.03.085
  [nucl-ex/0501009].
  %%CITATION = doi:10.1016/j.nuclphysa.2005.03.085;%%
  %2819 citations counted in INSPIRE as of 28 Dec 2018


%\cite{Adcox:2004mh}
\bibitem{Adcox:2004mh} 
  K.~Adcox {\it et al.} [PHENIX Collaboration],
  %``Formation of dense partonic matter in relativistic nucleus-nucleus collisions at RHIC: Experimental evaluation by the PHENIX collaboration,''
  Nucl.\ Phys.\ A {\bf 757}, 184 (2005)
  doi:10.1016/j.nuclphysa.2005.03.086
  [nucl-ex/0410003].
  %%CITATION = doi:10.1016/j.nuclphysa.2005.03.086;%%
  %2622 citations counted in INSPIRE as of 28 Dec 2018


%\cite{Arsene:2004fa}
\bibitem{Arsene:2004fa} 
  I.~Arsene {\it et al.} [BRAHMS Collaboration],
  %``Quark gluon plasma and color glass condensate at RHIC? The Perspective from the BRAHMS experiment,''
  Nucl.\ Phys.\ A {\bf 757}, 1 (2005)
  doi:10.1016/j.nuclphysa.2005.02.130
  [nucl-ex/0410020].
  %%CITATION = doi:10.1016/j.nuclphysa.2005.02.130;%%
  %1992 citations counted in INSPIRE as of 28 Dec 2018


%\cite{Back:2004je}
\bibitem{Back:2004je} 
  B.~B.~Back {\it et al.},
  %``The PHOBOS perspective on discoveries at RHIC,''
  Nucl.\ Phys.\ A {\bf 757}, 28 (2005)
  doi:10.1016/j.nuclphysa.2005.03.084
  [nucl-ex/0410022].
  %%CITATION = doi:10.1016/j.nuclphysa.2005.03.084;%%
  %2027 citations counted in INSPIRE as of 28 Dec 2018

%\cite{Yin:2013zea}
\bibitem{Yin:2013zea} 
  Z.~B.~Yin [ALICE Collaboration],
  %``Elliptic flow of strange and multi-strange particles in Pb-Pb collisions at $\sqrt{s_{NN}}$ = 2.76-TeV measured with ALICE,''
  Acta Phys.\ Polon.\ Supp.\  {\bf 6}, 479 (2013).
  doi:10.5506/APhysPolBSupp.6.479
  %%CITATION = doi:10.5506/APhysPolBSupp.6.479;%%
  %4 citations counted in INSPIRE as of 28 Dec 2018

%\cite{Aad:2010bu}
\bibitem{Aad:2010bu} 
  G.~Aad {\it et al.} [ATLAS Collaboration],
  %``Observation of a Centrality-Dependent Dijet Asymmetry in Lead-Lead Collisions at $\sqrt{s_{NN}}=2.77$ TeV with the ATLAS Detector at the LHC,''
  Phys.\ Rev.\ Lett.\  {\bf 105}, 252303 (2010)
  doi:10.1103/PhysRevLett.105.252303
  [arXiv:1011.6182 [hep-ex]].
  %%CITATION = doi:10.1103/PhysRevLett.105.252303;%%
  %716 citations counted in INSPIRE as of 28 Dec 2018


%\cite{Chatrchyan:2011sx}
\bibitem{Chatrchyan:2011sx} 
  S.~Chatrchyan {\it et al.} [CMS Collaboration],
  %``Observation and studies of jet quenching in PbPb collisions at nucleon-nucleon center-of-mass energy = 2.76 TeV,''
  Phys.\ Rev.\ C {\bf 84}, 024906 (2011)
  doi:10.1103/PhysRevC.84.024906
  [arXiv:1102.1957 [nucl-ex]].
  %%CITATION = doi:10.1103/PhysRevC.84.024906;%%
  %670 citations counted in INSPIRE as of 28 Dec 2018

%\cite{Burke:2013yra}
\bibitem{Burke:2013yra} 
  K.~M.~Burke {\it et al.} [JET Collaboration],
  %``Extracting the jet transport coefficient from jet quenching in high-energy heavy-ion collisions,''
  Phys.\ Rev.\ C {\bf 90}, no. 1, 014909 (2014)
  doi:10.1103/PhysRevC.90.014909
  [arXiv:1312.5003 [nucl-th]].
  %%CITATION = doi:10.1103/PhysRevC.90.014909;%%
  %204 citations counted in INSPIRE as of 28 Dec 2018

%\cite{DEramo:2010wup}
\bibitem{DEramo:2010wup} 
  F.~D'Eramo, H.~Liu and K.~Rajagopal,
  %``Transverse Momentum Broadening and the Jet Quenching Parameter, Redux,''
  Phys.\ Rev.\ D {\bf 84}, 065015 (2011)
  doi:10.1103/PhysRevD.84.065015
  [arXiv:1006.1367 [hep-ph]].
  %%CITATION = doi:10.1103/PhysRevD.84.065015;%%
  %101 citations counted in INSPIRE as of 28 Dec 2018


%\cite{Rapp:2009my}
\bibitem{Rapp:2009my}
R.~Rapp and H.~van Hees,
%``Heavy Quarks in the Quark-Gluon Plasma,''
doi:10.1142/9789814293297\_0003
[arXiv:0903.1096 [hep-ph]].
%175 citations counted in INSPIRE as of 05 Aug 2023


%\cite{Dunkel:2008ngc}
\bibitem{Dunkel:2008ngc}
J.~Dunkel and P.~H\"anggi,
%``Relativistic Brownian motion,''
Phys. Rept. \textbf{471}, 1-73 (2009)
doi:10.1016/j.physrep.2008.12.001
[arXiv:0812.1996 [cond-mat.stat-mech]].
%78 citations counted in INSPIRE as of 05 Aug 2023

%%%%%%%%%%%%%%%%%%%%%%%%%%%%%%%%%%%%%%%%%%%%%  Drag Force refs

%\cite{Domurcukgul:2021qfe}
\bibitem{Domurcukgul:2021qfe}
T.~Domurcukgul and R.~Morad,
%``Holographic drag force in non-conformal plasma,''
Eur. Phys. J. C \textbf{82} (2022) no.4, 304
doi:10.1140/epjc/s10052-022-10252-w
[arXiv:2108.10853 [hep-th]].
%0 citations counted in INSPIRE as of 16 Aug 2023

%\cite{Mes:2020vgy}
\bibitem{Mes:2020vgy}
A.~K.~Mes, R.~W.~Moerman, J.~P.~Shock and W.~A.~Horowitz,
%``Strongly coupled heavy and light quark thermal motion from AdS/CFT,''
Annals Phys. \textbf{436} (2022), 168675
doi:10.1016/j.aop.2021.168675
[arXiv:2008.09196 [hep-th]].
%2 citations counted in INSPIRE as of 16 Aug 2023

%\cite{Caceres:2006dj}
\bibitem{Caceres:2006dj}
E.~Caceres and A.~Guijosa,
%``Drag force in charged N=4 SYM plasma,''
JHEP \textbf{11}, 077 (2006)
doi:10.1088/1126-6708/2006/11/077
[arXiv:hep-th/0605235 [hep-th]].
%95 citations counted in INSPIRE as of 05 Aug 2023

%\cite{Chakraborty:2014kfa}
\bibitem{Chakraborty:2014kfa}
S.~Chakraborty and N.~Haque,
%``Drag force in strongly coupled, anisotropic plasma at finite chemical potential,''
JHEP \textbf{12}, 175 (2014)
doi:10.1007/JHEP12(2014)175
[arXiv:1410.7040 [hep-th]].
%15 citations counted in INSPIRE as of 05 Aug 2023

%\cite{Cheng:2014fza}
\bibitem{Cheng:2014fza}
L.~Cheng, X.~H.~Ge and S.~Y.~Wu,
%``Drag force of Anisotropic plasma at finite $U(1)$ chemical potential,''
Eur. Phys. J. C \textbf{76}, no.5, 256 (2016)
doi:10.1140/epjc/s10052-016-4096-7
[arXiv:1412.8433 [hep-th]].
%19 citations counted in INSPIRE as of 05 Aug 2023

%\cite{Fadafan:2008gb}
\bibitem{Fadafan:2008gb}
K.~B.~Fadafan,
%``R**2 curvature-squared corrections on drag force,''
JHEP \textbf{12}, 051 (2008)
doi:10.1088/1126-6708/2008/12/051
[arXiv:0803.2777 [hep-th]].
%55 citations counted in INSPIRE as of 05 Aug 2023

%\cite{Matsuo:2006ws}
\bibitem{Matsuo:2006ws}
T.~Matsuo, D.~Tomino and W.~Y.~Wen,
%``Drag force in SYM plasma with B field from AdS/CFT,''
JHEP \textbf{10}, 055 (2006)
doi:10.1088/1126-6708/2006/10/055
[arXiv:hep-th/0607178 [hep-th]].
%66 citations counted in INSPIRE as of 05 Aug 2023

%\cite{Talavera:2006tj}
\bibitem{Talavera:2006tj}
P.~Talavera,
%``Drag force in a string model dual to large-N QCD,''
JHEP \textbf{01}, 086 (2007)
doi:10.1088/1126-6708/2007/01/086
[arXiv:hep-th/0610179 [hep-th]].
%34 citations counted in INSPIRE as of 05 Aug 2023

%\cite{Zhang:2018mqt}
\bibitem{Zhang:2018mqt}
Z.~q.~Zhang, K.~Ma and D.~f.~Hou,
%``Drag force in strongly coupled supersymmetric Yang\textendash{}Mills plasma in a magnetic field,''
J. Phys. G \textbf{45}, no.2, 025003 (2018)
doi:10.1088/1361-6471/aaa097
[arXiv:1802.01912 [hep-th]].
%16 citations counted in INSPIRE as of 05 Aug 2023

%\cite{Andreev:2018emc}
\bibitem{Andreev:2018emc}
O.~Andreev,
%``Drag force on heavy diquarks and gauge/string duality,''
Phys. Rev. D \textbf{98}, no.6, 066007 (2018)
doi:10.1103/PhysRevD.98.066007
[arXiv:1804.09529 [hep-ph]].
%5 citations counted in INSPIRE as of 05 Aug 2023

%\cite{Sadeghi:2009mp}
\bibitem{Sadeghi:2009mp}
J.~Sadeghi, M.~R.~Setare, B.~Pourhassan and S.~Hashmatian,
%``Drag Force of Moving Quark in STU Background,''
Eur. Phys. J. C \textbf{61}, 527-533 (2009)
doi:10.1140/epjc/s10052-009-1011-5
[arXiv:0901.0217 [hep-th]].
%54 citations counted in INSPIRE as of 05 Aug 2023

%\cite{Gubser:2006qh}
\bibitem{Gubser:2006qh}
S.~S.~Gubser,
%``Comparing the drag force on heavy quarks in N=4 super-Yang-Mills theory and QCD,''
Phys. Rev. D \textbf{76}, 126003 (2007)
doi:10.1103/PhysRevD.76.126003
[arXiv:hep-th/0611272 [hep-th]].
%156 citations counted in INSPIRE as of 05 Aug 2023


%\cite{Herzog:2006gh}
\bibitem{Herzog:2006gh}
C.~P.~Herzog, A.~Karch, P.~Kovtun, C.~Kozcaz and L.~G.~Yaffe,
%``Energy loss of a heavy quark moving through N=4 supersymmetric Yang-Mills plasma,''
JHEP \textbf{07}, 013 (2006)
doi:10.1088/1126-6708/2006/07/013
[arXiv:hep-th/0605158 [hep-th]].
%745 citations counted in INSPIRE as of 05 Aug 2023

%\cite{Chernicoff:2012iq}
\bibitem{Chernicoff:2012iq}
M.~Chernicoff, D.~Fernandez, D.~Mateos and D.~Trancanelli,
%``Drag force in a strongly coupled anisotropic plasma,''
JHEP \textbf{08}, 100 (2012)
doi:10.1007/JHEP08(2012)100
[arXiv:1202.3696 [hep-th]].
%78 citations counted in INSPIRE as of 05 Aug 2023

%\cite{NataAtmaja:2010hd}
\bibitem{NataAtmaja:2010hd}
A.~Nata Atmaja and K.~Schalm,
%``Anisotropic Drag Force from 4D Kerr-AdS Black Holes,''
JHEP \textbf{04}, 070 (2011)
doi:10.1007/JHEP04(2011)070
[arXiv:1012.3800 [hep-th]].
%59 citations counted in INSPIRE as of 05 Aug 2023

%\cite{Panigrahi:2010cm}
\bibitem{Panigrahi:2010cm}
K.~L.~Panigrahi and S.~Roy,
%``Drag force in a hot non-relativistic, non-commutative Yang-Mills plasma,''
JHEP \textbf{04}, 003 (2010)
doi:10.1007/JHEP04(2010)003
[arXiv:1001.2904 [hep-th]].
%20 citations counted in INSPIRE as of 05 Aug 2023

%\cite{Xiong:2019wik}
\bibitem{Xiong:2019wik}
Y.~Xiong, X.~Tang and Z.~Luo,
%``Drag force on heavy quarks from holographic QCD,''
Chin. Phys. C \textbf{43}, no.11, 113103 (2019)
doi:10.1088/1674-1137/43/11/113103
[arXiv:1909.00928 [hep-ph]].
%6 citations counted in INSPIRE as of 05 Aug 2023

%\cite{Gursoy:2009kk}
\bibitem{Gursoy:2009kk}
U.~Gursoy, E.~Kiritsis, G.~Michalogiorgakis and F.~Nitti,
%``Thermal Transport and Drag Force in Improved Holographic QCD,''
JHEP \textbf{12}, 056 (2009)
doi:10.1088/1126-6708/2009/12/056
[arXiv:0906.1890 [hep-ph]].
%124 citations counted in INSPIRE as of 05 Aug 2023

%\cite{Gubser:2006bz}
\bibitem{Gubser:2006bz} 
  S.~S.~Gubser,
  %``Drag force in AdS/CFT,''
  Phys.\ Rev.\ D {\bf 74}, 126005 (2006)
  doi:10.1103/PhysRevD.74.126005
  [hep-th/0605182].
  %%CITATION = doi:10.1103/PhysRevD.74.126005;%%
  %563 citations counted in INSPIRE as of 28 Dec 2018

%\cite{Giataganas:2013zaa}
\bibitem{Giataganas:2013zaa}
D.~Giataganas and H.~Soltanpanahi,
%``Heavy Quark Diffusion in Strongly Coupled Anisotropic Plasmas,''
JHEP \textbf{06}, 047 (2014)
doi:10.1007/JHEP06(2014)047
[arXiv:1312.7474 [hep-th]].
%48 citations counted in INSPIRE as of 08 Dec 2023

%\cite{Zhu:2021nbl}
\bibitem{Zhu:2021nbl}
Z.~R.~Zhu, J.~X.~Chen, X.~M.~Liu and D.~Hou,
%``Thermodynamics and energy loss in D dimensions from holographic QCD model,''
Eur. Phys. J. C \textbf{82}, no.6, 560 (2022)
doi:10.1140/epjc/s10052-022-10433-7
[arXiv:2109.02366 [hep-ph]].
%2 citations counted in INSPIRE as of 28 Sep 2023

%%%%%%%%%%%%%%%%%%%%%%%%%%%%%%%%%%%%%%%%%%%%%% Langevin refs

%\cite{Gubser:2006nz}
\bibitem{Gubser:2006nz}
S.~S.~Gubser,
%``Momentum fluctuations of heavy quarks in the gauge-string duality,''
Nucl. Phys. B \textbf{790}, 175-199 (2008)
doi:10.1016/j.nuclphysb.2007.09.017
[arXiv:hep-th/0612143 [hep-th]].
%223 citations counted in INSPIRE as of 05 Aug 2023


%\cite{Casalderrey-Solana:2006fio}
\bibitem{Casalderrey-Solana:2006fio}
J.~Casalderrey-Solana and D.~Teaney,
%``Heavy quark diffusion in strongly coupled N=4 Yang-Mills,''
Phys. Rev. D \textbf{74}, 085012 (2006)
doi:10.1103/PhysRevD.74.085012
[arXiv:hep-ph/0605199 [hep-ph]].
%477 citations counted in INSPIRE as of 05 Aug 2023

%\cite{Casalderrey-Solana:2007ahi}
\bibitem{Casalderrey-Solana:2007ahi}
J.~Casalderrey-Solana and D.~Teaney,
%``Transverse Momentum Broadening of a Fast Quark in a N=4 Yang Mills Plasma,''
JHEP \textbf{04}, 039 (2007)
doi:10.1088/1126-6708/2007/04/039
[arXiv:hep-th/0701123 [hep-th]].
%214 citations counted in INSPIRE as of 05 Aug 2023

%\cite{deBoer:2008gu}
\bibitem{deBoer:2008gu}
J.~de Boer, V.~E.~Hubeny, M.~Rangamani and M.~Shigemori,
%``Brownian motion in AdS/CFT,''
JHEP \textbf{07}, 094 (2009)
doi:10.1088/1126-6708/2009/07/094
[arXiv:0812.5112 [hep-th]].
%125 citations counted in INSPIRE as of 05 Aug 2023

%\cite{Son:2009vu}
\bibitem{Son:2009vu}
D.~T.~Son and D.~Teaney,
%``Thermal Noise and Stochastic Strings in AdS/CFT,''
JHEP \textbf{07}, 021 (2009)
doi:10.1088/1126-6708/2009/07/021
[arXiv:0901.2338 [hep-th]].
%119 citations counted in INSPIRE as of 05 Aug 2023

%\cite{Giecold:2009cg}
\bibitem{Giecold:2009cg}
G.~C.~Giecold, E.~Iancu and A.~H.~Mueller,
%``Stochastic trailing string and Langevin dynamics from AdS/CFT,''
JHEP \textbf{07}, 033 (2009)
doi:10.1088/1126-6708/2009/07/033
[arXiv:0903.1840 [hep-th]].
%69 citations counted in INSPIRE as of 05 Aug 2023


%\cite{Iqbal:2008by}
\bibitem{Iqbal:2008by}
N.~Iqbal and H.~Liu,
%``Universality of the hydrodynamic limit in AdS/CFT and the membrane paradigm,''
Phys. Rev. D \textbf{79}, 025023 (2009)
doi:10.1103/PhysRevD.79.025023
[arXiv:0809.3808 [hep-th]].
%557 citations counted in INSPIRE as of 27 Sep 2023

%\cite{Gursoy:2010aa}
\bibitem{Gursoy:2010aa}
U.~Gursoy, E.~Kiritsis, L.~Mazzanti and F.~Nitti,
%``Langevin diffusion of heavy quarks in non-conformal holographic backgrounds,''
JHEP \textbf{12}, 088 (2010)
doi:10.1007/JHEP12(2010)088
[arXiv:1006.3261 [hep-th]].
%65 citations counted in INSPIRE as of 05 Aug 2023

%\cite{Giataganas:2013hwa}
\bibitem{Giataganas:2013hwa}
D.~Giataganas and H.~Soltanpanahi,
%``Universal Properties of the Langevin Diffusion Coefficients,''
Phys. Rev. D \textbf{89}, no.2, 026011 (2014)
doi:10.1103/PhysRevD.89.026011
[arXiv:1310.6725 [hep-th]].
%54 citations counted in INSPIRE as of 05 Aug 2023

%\cite{Kiritsis:2013iba}
\bibitem{Kiritsis:2013iba}
E.~Kiritsis, L.~Mazzanti and F.~Nitti,
%``The confining trailing string,''
JHEP \textbf{02}, 081 (2014)
doi:10.1007/JHEP02(2014)081
[arXiv:1311.2611 [hep-th]].
%7 citations counted in INSPIRE as of 05 Aug 2023

%\cite{Mykhaylova:2020pfk}
\bibitem{Mykhaylova:2020pfk}
V.~Mykhaylova and C.~Sasaki,
%``Impact of quark quasiparticles on transport coefficients in hot QCD,''
Phys. Rev. D \textbf{103}, no.1, 014007 (2021)
doi:10.1103/PhysRevD.103.014007
[arXiv:2007.06846 [hep-ph]].
%14 citations counted in INSPIRE as of 05 Aug 2023

%\cite{Zhu:2020wds}
\bibitem{Zhu:2020wds}
X.~Zhu and Z.~Q.~Zhang,
%``Light quark energy loss in a soft-wall AdS/QCD model,''
Eur. Phys. J. A \textbf{57}, no.3, 96 (2021)
doi:10.1140/epja/s10050-021-00418-7
[arXiv:2011.00920 [nucl-th]].
%2 citations counted in INSPIRE as of 05 Aug 2023


%%%%%%%%%%%%%%%%%%%%%%%%%%%%%%%%%%%%%%%%%%%%%%
%%%%%%%%%%%%%%%%%%%%%%%%%%%%%%%%%%%%%%%%%%%%%%
%%%%%%%%%%%%%%%%%%%%%%%%%%%%%%%%%%%%%%%%%%%%%%


%\cite{Liu:2006ug}
\bibitem{Liu:2006ug} 
  H.~Liu, K.~Rajagopal and U.~A.~Wiedemann,
  %``Calculating the jet quenching parameter from AdS/CFT,''
  Phys.\ Rev.\ Lett.\  {\bf 97}, 182301 (2006)
  doi:10.1103/PhysRevLett.97.182301
  [hep-ph/0605178].
  %%CITATION = doi:10.1103/PhysRevLett.97.182301;%%
  %400 citations counted in INSPIRE as of 28 Dec 2018

  %\cite{Rougemont:2015wca}
\bibitem{Rougemont:2015wca} 
  R.~Rougemont, A.~Ficnar, S.~Finazzo and J.~Noronha,
  %``Energy loss, equilibration, and thermodynamics of a baryon rich strongly coupled quark-gluon plasma,''
  JHEP {\bf 1604}, 102 (2016)
  doi:10.1007/JHEP04(2016)102
  [arXiv:1507.06556 [hep-th]].
  %%CITATION = doi:10.1007/JHEP04(2016)102;%%
  %44 citations counted in INSPIRE as of 29 Dec 2018
  
 %\cite{Li:2014hja}
\bibitem{Li:2014hja} 
  D.~Li, J.~Liao and M.~Huang,
  %``Enhancement of jet quenching around phase transition: result from the dynamical holographic model,''
  Phys.\ Rev.\ D {\bf 89}, no. 12, 126006 (2014)
  doi:10.1103/PhysRevD.89.126006
  [arXiv:1401.2035 [hep-ph]].
  %%CITATION = doi:10.1103/PhysRevD.89.126006;%%
  %32 citations counted in INSPIRE as of 29 Dec 2018


%\cite{Brandhuber:1998bs}
\bibitem{Brandhuber:1998bs} 
  A.~Brandhuber, N.~Itzhaki, J.~Sonnenschein and S.~Yankielowicz,
  %``Wilson loops in the large N limit at finite temperature,''
  Phys.\ Lett.\ B {\bf 434}, 36 (1998)
  doi:10.1016/S0370-2693(98)00730-8
  [hep-th/9803137].
  %%CITATION = doi:10.1016/S0370-2693(98)00730-8;%%
  %297 citations counted in INSPIRE as of 28 Dec 2018


%\cite{Sonnenschein:1999if}
\bibitem{Sonnenschein:1999if} 
  J.~Sonnenschein,
  %``What does the string / gauge correspondence teach us about Wilson loops?,''
  hep-th/0003032.
  %%CITATION = HEP-TH/0003032;%%
  %89 citations counted in INSPIRE as of 28 Dec 2018

%\cite{Liu:2006he}
\bibitem{Liu:2006he} 
  H.~Liu, K.~Rajagopal and U.~A.~Wiedemann,
  %``Wilson loops in heavy ion collisions and their calculation in AdS/CFT,''
  JHEP {\bf 0703}, 066 (2007)
  doi:10.1088/1126-6708/2007/03/066
  [hep-ph/0612168].
  %%CITATION = doi:10.1088/1126-6708/2007/03/066;%%
  %223 citations counted in INSPIRE as of 28 Dec 2018

%\cite{Caceres:2006as}
\bibitem{Caceres:2006as} 
  E.~Caceres and A.~Guijosa,
  %``On Drag Forces and Jet Quenching in Strongly Coupled Plasmas,''
  JHEP {\bf 0612}, 068 (2006)
  doi:10.1088/1126-6708/2006/12/068
  [hep-th/0606134].
  %%CITATION = doi:10.1088/1126-6708/2006/12/068;%%
  %76 citations counted in INSPIRE as of 28 Dec 2018

%\cite{VazquezPoritz:2006ba}
\bibitem{VazquezPoritz:2006ba} 
  J.~F.~Vazquez-Poritz,
  %``Enhancing the jet quenching parameter from marginal deformations,''
  hep-th/0605296.
  %%CITATION = HEP-TH/0605296;%%
  %48 citations counted in INSPIRE as of 28 Dec 2018

%\cite{Nakano:2006js}
\bibitem{Nakano:2006js} 
  E.~Nakano, S.~Teraguchi and W.~Y.~Wen,
  %``Drag force, jet quenching, and AdS/QCD,''
  Phys.\ Rev.\ D {\bf 75}, 085016 (2007)
  doi:10.1103/PhysRevD.75.085016
  [hep-ph/0608274].
  %%CITATION = doi:10.1103/PhysRevD.75.085016;%%
  %59 citations counted in INSPIRE as of 28 Dec 2018

%\cite{Avramis:2006ip}
\bibitem{Avramis:2006ip} 
  S.~D.~Avramis and K.~Sfetsos,
  %``Supergravity and the jet quenching parameter in the presence of R-charge densities,''
  JHEP {\bf 0701}, 065 (2007)
  doi:10.1088/1126-6708/2007/01/065
  [hep-th/0606190].
  %%CITATION = doi:10.1088/1126-6708/2007/01/065;%%
  %71 citations counted in INSPIRE as of 28 Dec 2018

%\cite{Gao:2006uf}
\bibitem{Gao:2006uf} 
  Y.~h.~Gao, W.~s.~Xu and D.~f.~Zeng,
  %``Jet quenching parameters of Sakai-Sugimoto Model,''
  hep-th/0611217.
  %%CITATION = HEP-TH/0611217;%%
  %23 citations counted in INSPIRE as of 28 Dec 2018

%\cite{Armesto:2006zv}
\bibitem{Armesto:2006zv} 
  N.~Armesto, J.~D.~Edelstein and J.~Mas,
  %``Jet quenching at finite `t Hooft coupling and chemical potential from AdS/CFT,''
  JHEP {\bf 0609}, 039 (2006)
  doi:10.1088/1126-6708/2006/09/039
  [hep-ph/0606245].
  %%CITATION = doi:10.1088/1126-6708/2006/09/039;%%
  %108 citations counted in INSPIRE as of 28 Dec 2018

%\cite{Lin:2006au}
\bibitem{Lin:2006au} 
  F.~L.~Lin and T.~Matsuo,
  %``Jet Quenching Parameter in Medium with Chemical Potential from AdS/CFT,''
  Phys.\ Lett.\ B {\bf 641}, 45 (2006)
  doi:10.1016/j.physletb.2006.08.024
  [hep-th/0606136].
  %%CITATION = doi:10.1016/j.physletb.2006.08.024;%%
  %72 citations counted in INSPIRE as of 28 Dec 2018

%\cite{Sadeghi:2013dga}
\bibitem{Sadeghi:2013dga} 
  J.~Sadeghi and S.~Heshmatian,
  %``Jet Quenching Parameter with Hyperscaling Violation,''
  Eur.\ Phys.\ J.\ C {\bf 74}, 3032 (2014)
  doi:10.1140/epjc/s10052-014-3032-y
  [arXiv:1308.5991 [hep-th]].
  %%CITATION = doi:10.1140/epjc/s10052-014-3032-y;%%
  %7 citations counted in INSPIRE as of 28 Dec 2018

%\cite{Cai:2012eh}
\bibitem{Cai:2012eh} 
  R.~G.~Cai, S.~Chakrabortty, S.~He and L.~Li,
  %``Some aspects of QGP phase in a hQCD model,''
  JHEP {\bf 1302}, 068 (2013)
  doi:10.1007/JHEP02(2013)068
  [arXiv:1209.4512 [hep-th]].
  %%CITATION = doi:10.1007/JHEP02(2013)068;%%
  %20 citations counted in INSPIRE as of 28 Dec 2018

%\cite{Wang:2016noh}
\bibitem{Wang:2016noh} 
  L.~Wang and S.~Y.~Wu,
  %``Holographic study of the jet quenching parameter in anisotropic systems,''
  Eur.\ Phys.\ J.\ C {\bf 76}, no. 11, 587 (2016)
  doi:10.1140/epjc/s10052-016-4421-1
  [arXiv:1609.03665 [hep-th]].
  %%CITATION = doi:10.1140/epjc/s10052-016-4421-1;%%
  %3 citations counted in INSPIRE as of 28 Dec 2018

%\cite{DeWolfe:2009vs}
\bibitem{DeWolfe:2009vs} 
  O.~DeWolfe and C.~Rosen,
  %``Robustness of Sound Speed and Jet Quenching for Gauge/Gravity Models of Hot QCD,''
  JHEP {\bf 0907}, 022 (2009)
  doi:10.1088/1126-6708/2009/07/022
  [arXiv:0903.1458 [hep-th]].
  %%CITATION = doi:10.1088/1126-6708/2009/07/022;%%
  %17 citations counted in INSPIRE as of 28 Dec 2018

%\cite{Fadafan:2008uv}
\bibitem{Fadafan:2008uv} 
  K.~Bitaghsir Fadafan,
  %``Charge effect and finite 't Hooft coupling correction on drag force and Jet Quenching Parameter,''
  Eur.\ Phys.\ J.\ C {\bf 68}, 505 (2010)
  doi:10.1140/epjc/s10052-010-1375-6
  [arXiv:0809.1336 [hep-th]].
  %%CITATION = doi:10.1140/epjc/s10052-010-1375-6;%%
  %43 citations counted in INSPIRE as of 28 Dec 2018

%\cite{Horowitz:2017nbm}
\bibitem{Horowitz:2017nbm} 
  W.~A.~Horowitz,
  %``Time Dependent $\hat{q}$ from AdS/CFT,''
  Nucl.\ Part.\ Phys.\ Proc.\  {\bf 289-290}, 129 (2017).
  doi:10.1016/j.nuclphysbps.2017.05.026
  %%CITATION = doi:10.1016/j.nuclphysbps.2017.05.026;%%

%\cite{Horowitz:2015dta}
\bibitem{Horowitz:2015dta} 
  W.~A.~Horowitz,
  %``Fluctuating heavy quark energy loss in a strongly coupled quark-gluon plasma,''
  Phys.\ Rev.\ D {\bf 91}, no. 8, 085019 (2015)
  doi:10.1103/PhysRevD.91.085019
  [arXiv:1501.04693 [hep-ph]].
  %%CITATION = doi:10.1103/PhysRevD.91.085019;%%
  %26 citations counted in INSPIRE as of 28 Dec 2018

%\cite{Heshmatian:2018wlv}
\bibitem{Heshmatian:2018wlv}
S.~Heshmatian, R.~Morad and M.~Akbari,
%``Jet suppression in non-conformal plasma using AdS/CFT,''
JHEP \textbf{03}, 045 (2019)
doi:10.1007/JHEP03(2019)045
[arXiv:1812.09374 [hep-th]].
%5 citations counted in INSPIRE as of 05 Aug 2023

%\cite{Giataganas:2012zy}
\bibitem{Giataganas:2012zy}
D.~Giataganas,
%``Probing strongly coupled anisotropic plasma,''
JHEP \textbf{07}, 031 (2012)
doi:10.1007/JHEP07(2012)031
[arXiv:1202.4436 [hep-th]].
%147 citations counted in INSPIRE as of 08 Dec 2023

%\cite{Morad:2014xla}
\bibitem{Morad:2014xla} 
  R.~Morad and W.~A.~Horowitz,
  %``Strong-coupling Jet Energy Loss from AdS/CFT,''
  JHEP {\bf 1411}, 017 (2014)
  doi:10.1007/JHEP11(2014)017
  [arXiv:1409.7545 [hep-th]].
  %%CITATION = doi:10.1007/JHEP11(2014)017;%%
  %16 citations counted in INSPIRE as of 28 Dec 2018

%\cite{BitaghsirFadafan:2017tci}
\bibitem{BitaghsirFadafan:2017tci} 
  K.~Bitaghsir Fadafan and R.~Morad,
  %``Jets in a strongly coupled anisotropic plasma,''
  Eur.\ Phys.\ J.\ C {\bf 78}, no. 1, 16 (2018)
  doi:10.1140/epjc/s10052-018-5520-y
  [arXiv:1710.06417 [hep-th]].
  %%CITATION = doi:10.1140/epjc/s10052-018-5520-y;%%
  %4 citations counted in INSPIRE as of 28 Dec 2018

%\cite{Chesler:2008uy}
\bibitem{Chesler:2008uy} 
  P.~M.~Chesler, K.~Jensen, A.~Karch and L.~G.~Yaffe,
  %``Light quark energy loss in strongly-coupled N = 4 supersymmetric Yang-Mills plasma,''
  Phys.\ Rev.\ D {\bf 79}, 125015 (2009)
  doi:10.1103/PhysRevD.79.125015
  [arXiv:0810.1985 [hep-th]].
  %%CITATION = doi:10.1103/PhysRevD.79.125015;%%
  %134 citations counted in INSPIRE as of 28 Dec 2018

%\cite{Ficnar:2011yj}
\bibitem{Ficnar:2011yj} 
  A.~Ficnar, J.~Noronha and M.~Gyulassy,
  %``Jet Quenching in Non-Conformal Holography,''
  J.\ Phys.\ G {\bf 38}, 124176 (2011)
  doi:10.1088/0954-3899/38/12/124176
  [arXiv:1106.6303 [hep-ph]].
  %%CITATION = doi:10.1088/0954-3899/38/12/124176;%%
  %28 citations counted in INSPIRE as of 28 Dec 2018

%\cite{Chesler:2008wd}
\bibitem{Chesler:2008wd} 
  P.~M.~Chesler, K.~Jensen and A.~Karch,
  %``Jets in strongly-coupled N = 4 super Yang-Mills theory,''
  Phys.\ Rev.\ D {\bf 79}, 025021 (2009)
  doi:10.1103/PhysRevD.79.025021
  [arXiv:0804.3110 [hep-th]].
  %%CITATION = doi:10.1103/PhysRevD.79.025021;%%
  %71 citations counted in INSPIRE as of 28 Dec 2018

%\cite{Zakharov:1997uu}
\bibitem{Zakharov:1997uu} 
  B.~G.~Zakharov,
  %``Radiative energy loss of high-energy quarks in finite size nuclear matter and quark - gluon plasma,''
  JETP Lett.\  {\bf 65}, 615 (1997)
  doi:10.1134/1.567389
  [hep-ph/9704255].
  %%CITATION = doi:10.1134/1.567389;%%
  %523 citations counted in INSPIRE as of 28 Dec 2018

%\cite{Giataganas:2012zy}
\bibitem{Giataganas:2012zy}
D.~Giataganas,
%``Probing strongly coupled anisotropic plasma,''
JHEP \textbf{07}, 031 (2012)
doi:10.1007/JHEP07(2012)031
[arXiv:1202.4436 [hep-th]].
%147 citations counted in INSPIRE as of 08 Dec 2023

%\cite{Zhou:2022izh}
\bibitem{Zhou:2022izh}
Q.~Zhou and B.~W.~Zhang,
%``Holographic energy loss near critical temperature in an anisotropic background,''
Commun. Theor. Phys. \textbf{75}, no.10, 105301 (2023)
doi:10.1088/1572-9494/acea23
[arXiv:2211.14792 [hep-ph]].
%0 citations counted in INSPIRE as of 08 Dec 2023

%\cite{Arefeva:2020byn}
\bibitem{Arefeva:2020byn}
I.~Y.~Aref'eva, K.~Rannu and P.~Slepov,
%``Holographic anisotropic model for light quarks with confinement-deconfinement phase transition,''
JHEP \textbf{06}, 090 (2021)
doi:10.1007/JHEP06(2021)090
[arXiv:2009.05562 [hep-th]].
%27 citations counted in INSPIRE as of 08 Dec 2023

\bibitem{heshmatian2023thermal}  
Sara Heshmatian and Alexander Trounev,
% ``Thermal quench of a dynamical QCD model in an external electric field,''
 [arXiv:2308.11856 [hep-th]].


\end{thebibliography}
\end{document}